\documentclass[]{aa} 
\usepackage{graphicx}
\usepackage{txfonts}
\begin{document}
%
\def\hi {H\,{\sc i}}
\def\hii {H\,{\sc ii}}
\def\water {H$_2$O}
\def\meth {CH$_{3}$OH}
\def\dg{$^{\circ}$}
\def\kms{km\,s$^{-1}$}
\def\ms{m\,s$^{-1}$}
\def\jyb{Jy\,beam$^{-1}$}
\def\mjyb{mJy\,beam$^{-1}$}
\def\solmass {\hbox{M$_{\odot}$}}
\def\solum {\hbox{L$_{\odot}$}} 
\def\d {$^{\circ}$}
\def\n {$n_{\rm{H_{2}}}$}
\def\kmsg{km\,s$^{-1}$\,G$^{-1}$}
\def\tbo {$T_{\rm{b}}\Delta\Omega$}
\def\tb {$T_{\rm{b}}$}
\def\om{$\Delta\Omega$}
\def\dvi {$\Delta V_{\rm{i}}$}
\def\dvz {$\Delta V_{\rm{Z}}$}
\def\code {full radiative transfer method code}
\title{EVN observations of 6.7~GHz methanol maser polarization in massive star-forming regions II. First statistical results.}

\author{G.\ Surcis  \inst{1}
  \and 
  W.H.T. \ Vlemmings \inst{2}
 \and
  H.J. \ van Langevelde \inst{1,3}
  \and
  B. \ Hutawarakorn Kramer \inst{4,5}
  \and
  L.H. Quiroga-Nu\~{n}ez \inst{6}
  }

\institute{Joint Institute for VLBI in Europe, Postbus 2, 7990 AA Dwingeloo, The Netherlands
 \email{surcis@jive.nl}
 \and
 Chalmers University of Technology, Onsala Space Observatory, SE-439 92 Onsala, Sweden
 \and
 Sterrewacht Leiden, Leiden University, Postbus 9513, 2300 RA Leiden, The Netherlands
 \and
 Max-Planck Institut f\"{u}r Radioastronomie, Auf dem H\"{u}gel 69, 53121 Bonn, Germany
 \and
 National Astronomical Research Institute of Thailand, Ministry of Science and Technology, Rama VI Rd., Bangkok 10400, Thailand
 \and
 Planetario de Bogot\'{a}, IDARTES, Calle 26 B No.5 - 93, Bogot\'{a}, Colombia
  }

\date{Received ; accepted}
\abstract
{Magnetic fields have only recently been included in theoretical simulations of high-mass star formation. The simulations show 
that magnetic fields play an
important role in the formation and dynamics of molecular outflows. Masers, in particular 6.7-GHz \meth ~masers, are the best probes
of the magnetic field morphologies around massive young stellar objects on the smallest scales of 10-100~AU.}
{Providing new observational measurements of the morphology of magnetic fields around massive young stellar objects by using 6.7-GHz 
\meth ~maser emission
is very important for setting constraints on the numerical simulations of massive star formation.} 
{This paper focuses on 4 massive young stellar objects, IRAS\,06058+2138-NIRS~1, IRAS\,22272+6358A, S255-IR, and S231, which complement our 
previous 2012 sample (the first EVN group). From all these sources, 
molecular outflows have been detected in the past. Seven of the European VLBI Network antennas were used to measure the linear polarization and 
Zeeman-splitting of the 6.7-GHz \meth ~masers in the star-forming regions in this second EVN group.}
{We detected a total of 128 \meth ~masing cloudlets. Fractional linear polarization (0.8~\% -- 11.3~\%) was detected towards 18\% of the 
\meth ~masers in our sample. The linear 
polarization  vectors are well ordered in all the massive young stellar objects. We measured significant Zeeman-splitting in IRAS\,06058+2138-NIRS~1 
(\dvz$=3.8\pm0.6$~\ms) and S255-IR (\dvz$=3.2\pm0.7$~\ms).}
{By considering the 20 massive young stellar objects towards which the morphology of magnetic fields was determined by observing 6.7-GHz \meth ~masers
 in both hemispheres, we find no evident correlation between the linear distributions of \meth ~masers and the outflows or the linear polarization
vectors. On the other hand, we present first statistical evidence that the magnetic field (on scales 10-100~AU) is primarily 
oriented along the large-scale outflow direction. Moreover, we empirically find that the linear polarization fraction of unsaturated \meth ~masers
is $P_{\rm{l}}<4.5\%$.}
\titlerunning{Magnetic field and outflows: first statistical results.}
\authorrunning{Surcis et al.}

\maketitle
\section{Introduction}
\label{intro}
Despite the likely importance of magnetic fields in the formation of low-mass stars (e.g., Matsumoto \& Tomisaka \cite{mat04}, 
McKee \& Ostriker \cite{mck07}), there are still only a few observations around massive young stellar objects (YSOs)
(e.g., Crutcher \cite{cru05}, Vlemmings et al. \cite{vle06}, Girart et al. \cite{gir09}), and theoretical
simulations of massive star formation have only recently included them (e.g., Banerjee \& Pudritz \cite{ban07}, Peters et al. \cite{pet11}, 
Seifried et al. \cite{sei12}). Since massive stars are fully radiative, they are, with few exceptions (e.g., Donati et al.
\cite{don06}, Alecian et al. \cite{ale12}), not expected to have significant magnetic fields. Consequently, it has been thought that the magnetic 
fields do not play any role in their formation. However, the detection of molecular outflows in massive star-forming regions and their 
necessary inclusion in the main models (e.g., McKee \& Tan \cite{mck03}, Bonnell et al. \cite{bon04}) makes the presence of magnetic fields
 during the formation of high-mass stars unavoidable. These fields are very likely the magnetic field frozen into the collapsing protostellar 
envelope.\\
\indent The formation of outflows has been observed in all simulations that include magnetic fields.
Banerjee \& Pudritz (\cite{ban07}) indicate
that magnetic fields coupled to the prestellar disks could be the possible driving power for early outflows, which can be understood 
in terms of a growing magnetic tower. By producing cavities through which radiation pressure can be released, these early outflows
 reduce the limitations on the final mass of massive stars imposed by simply considering the gravitational collapse. 
The outflows seem to be relatively fast and well-collimated for low and intermediate magnetic intensities ($\mu$\footnote{To evaluate 
the importance of the magnetic fields it is fundamental to compare the mass to magnetic flux ratio ($\rm{M/\Phi}$) to the critical value of this 
ratio, $(M/\Phi)_{\rm{crit}}\approx0.12/\sqrt{G}$, that is $\mu=(M/\Phi)/(M/\Phi)_{\rm{crit}}$. The stronger the magnetic field is, the lower $\mu$ is. 
If $\mu<1$, the magnetic field can prevent the collapse (Mouschovias \& Spitzer \cite{mou76}).}$=30-120$), and more slowly and poorly 
collimated for stronger fields ($\mu\sim5$; Hennebelle 
et al. \cite{hen11}, Seifried et al. \cite{sei12}). Furthermore, Seifried et al. (\cite{sei12}) show that magneto-centrifugally driven
 outflows 
consist of two regimes. In the first regime close to the disk and the rotation axis, acceleration is dominated by the centrifugal 
force; that is gas gets flung outwards along the poloidal magnetic field lines, whereas in the second regime farther away from the 
disk the toroidal magnetic field 
starts to dominate the acceleration. They also suggest that, for strong magnetic fields, the poorly collimated outflows are typical of 
the very early stage of massive star formation, and the collimation will subsequently increase because the launching of a 
well-collimated, fast jet 
overtakes the slowly expanding outflow. Peters et al. (\cite{pet11}) suggest two effects that tend to weaken and broaden the 
outflows. The first one comes from the disruption of the velocity coherence due to the gravitational fragmentation of the accretion flow. 
Second, 
the thermal pressure of ionized gas is higher than the magnetic pressure, so it is dynamically dominant within the \hii ~region.
Because the magnetically driven jets can survive until gravitational fragmentation disrupts uniform rotation, they therefore 
proposed the ionization feedback as a better driving source of the observed uncollimated outflows rather than the magnetic field.\\
\indent Besides contributing to the formation of outflows, the simulations show that magnetic fields prevent fragmentation, reduce angular 
momentum via magnetic braking, and, marginally, influences the accretion rate (Banerjee \& Pudritz \cite{ban07}, Peters et al. \cite{pet11}, 
Hennebelle et al. \cite{hen11}, Seifried et al. \cite{sei11}). The magnetic fields also play a significant role in the evolution of the 
circumstellar disk. While for weak magnetic fields, $\mu>10$, Keplerian disks with sizes of a few 100~AU are easily formed, for strong magnetic
fields ($\mu<10$) the Keplerian disks are formed only if a turbulent velocity field is introduced in the simulations (Seifried et al. \cite{sei11, 
sei12b}). Finally, magnetic fields determine also the size of \hii ~regions that in the presence of strong magnetic field are generally smaller
than without magnetic field (Peters et al. \cite{pet11}).\\
\indent Therefore, new measurements of the orientation and strength of magnetic fields at milliarcsecond (mas) resolution close 
to the massive YSOs are fundamental for providing new input for numerical simulations of massive star formation. Over the last years,
the high importance of using masers as probes of magnetic fields on the smallest scales (10-100~AU) has been proven
 (e.g., Vlemmings et al. \cite{vle06,vle10}, Surcis et al. \cite{sur11a, sur11b, sur12}). In particular, the 6.7-GHz \meth ~masers, which are 
among the most abundant maser species in massive star-forming regions, are playing a crucial role in determining the magnetic field morphology (e.g., 
Surcis et al. \cite{sur09, sur11b, sur12}). Magnetic fields have mainly been detected along outflows and in a few cases
on surfaces of disk/tori (Vlemmings et al. \cite{vle10}; Surcis et al. \cite{sur09, sur11b, sur12}). Moreover, 6.7-GHz \meth ~masers are also ideal 
for measuring the 
Zeeman-splitting even though the exact proportionality between the measured splitting and the magnetic field strength is still 
uncertain
(Vlemmings et al. \cite{vle11}). Therefore, enlarging the number of massive YSOs towards 
which observations in full polarization of 6.7-GHz \meth ~maser are made is of high importance. Here we show the results of our
 second EVN group
composed of 4 massive star-forming regions, which are described in details in Sect.~\ref{SEVNG}. The observations with the data reduction 
details are described in Sect.~\ref{obssect}, while the codes used for our analysis are introduced in Sect.~\ref{analysis}.
The results, which are presented in Sect.~\ref{res}, 
are discussed in Sect.~\ref{discussion}, where we statistically analyze the sample composed of all the massive YSOs towards
 which the morphology of magnetic fields was determined by observing the 6.7-GHz \meth ~masers. 
\section{The second EVN group}
\label{SEVNG}
We selected a subgroup of five massive star-forming regions among the northern
hemisphere sources observed with the Effelsberg 100-m telescope (Vlemmings \cite{vle08}, Vlemmings et al. \cite{vle11}). Hereafter, we
refer to this group as the second EVN group. Here, we present four of them while the results of IRAS20126+4104 will be discussed in 
a subsequent paper along with the results from
 22-GHz \water ~maser polarization observations. The sources were selected based on their high peak flux density to allow potential
detection of Zeeman-splitting as well as the presence of molecular outflows.
\subsection{IRAS\,06058+2138-NIRS~1}
IRAS\,06058+2138 (better known as S252 or AFGL5180) is a near-infrared (NIR) cluster of YSOs at a parallax distance 
of $1.76\pm0.11$~kpc (Oh et al. \cite{oh10}).
Three massive clumps were identified in the region by Saito et al. (\cite{sai07}). The most massive one named MCS~B coincides with
the MM1 clump ($\rm{M}\approx50$~\solmass) detected at 1.2-mm by Minier et al. (\cite{min05}). The massive YSO NIRS~1 
($V_{\rm{lsr}}^{\rm{C^{18}O}}=+3.9$~\kms, Saito et al.  \cite{sai07}; Tamura et al. \cite{tam91}) is associated with MCS~B and 
is also the 6.7-GHz \meth ~maser site (Minier et al. \cite{min00}, Xu et al. \cite{xu09}).
The \meth ~masers show a linear distribution of 120~mas with a roughly linear velocity gradient along it, although a few masers 
following a different velocity distribution (Minier et al. \cite{min00}). A CO-outflow ($\rm{PA_{out}}=130$\d, Snell et al. \cite{sne88}; Wu et al. 
\cite{wu10}) is
 centered and perpendicular to the linear distribution of the \meth ~masers (Wu et al. \cite{wu10}). Wu et al. (\cite{wu10}) measured
a velocity range for the blue-shifted and red-shifted lobes of $-10$~\kms$<V_{\rm{IRAS06058}}^{\rm{blue}}<-2$~\kms ~and 
$+8$~\kms$<V_{\rm{IRAS06058}}^{\rm{red}}<+20$~\kms, respectively. A Zeeman-splitting of the 6.7-GHz \meth ~masers of 
$\Delta V_{\rm{Z}}=-0.49\pm0.15$~\ms ~was measured by Vlemmings et al. (\cite{vle11}) with the Effelsberg 100-m telescope.
\subsection{IRAS\,22272+6358A}
IRAS\,22272+6358A is a YSO deeply embedded in the bright-rimmed cloud L1206 (Sugitani et al. \cite{sug91}). 
L1206 is located in the Local Arm at a parallax distance of 0.776$^{+0.104}_{-0.083}$~kpc (Rygl et al. \cite{ryg10}). No radio continuum emission
at 2~cm and 6~cm has been detected towards IRAS\,22272+6358A (Wilking et al. \cite{wil89}; McCutcheon et al. \cite{mcc91}) indicating that an 
\hii ~region has not been formed yet. Considering also the low color temperature $T(60\rm{\mu m}/100\rm{\mu m})\approx38$~K measured by Casoli
 et al. (\cite{cas86}), IRAS\,22272+6358A is a massive YSO at a very young phase (14.2~\solmass ~and $V_{\rm{lsr}}\backsimeq-11$~\kms; Beltr\'{a}n et al. 
\cite{bel06}). Beltr\'{a}n et al. (\cite{bel06}) also observed a CO-outlow ($\rm{PA_{out}}=140$\d) centered at the position of 
IRAS\,22272+6358A with velocity ranges $-19.5$~\kms$<V_{\rm{IRAS22272}}^{\rm{blue}}<-13.5$~\kms ~and 
$-8.5$~\kms$<V_{\rm{IRAS22272}}^{\rm{red}}<-2.5$~\kms ~of the blue-shifted and red-shifted lobes, respectively.\\
\indent At mas resolution two 6.7-GHz \meth ~maser groups were detected, which are separated by $\sim$100~mas and associated
with IRAS\,22272+6358A. A third weak group of \meth ~masers was also found northeast of the other two (Rygl et al. \cite{ryg10}). Vlemmings et
 al. (\cite{vle11}) measured a small Zeeman-splitting of the 6.7-GHz \meth ~masers of $\Delta V_{\rm{Z}}=0.53\pm0.15$~\ms. 
\subsection{S255-IR}
S255-IR ($V_{\rm{lsr, CO}}^{\rm{S255}}=+5.2$~\kms, Wang et al. \cite{wan11}) is a famous star-forming region located between the two
 \hii ~regions S255 and S257 at a parallax distance of 1.59$^{+0.07}_{-0.06}$~kpc
 (Rygl et al. \cite{ryg10}). Minier et al. (\cite{min05}) identified, with SCUBA, three mm sources in a dusty filament elongated north-south: MM1, 
MM2, and MM3. MM2, which contains three ultra-compact \hii ~regions (UC\hii), is associated with a NIR cluster of YSOs and appears to be 
the most evolved of the three regions (Minier et al. \cite{min05}). Wang et al. (\cite{wan11}) detected three 1.3~mm continuum peaks towards MM2
 and the stronger one named as SMA\,1 coincides with the near-infrared source NIRS\,3 (Tamura et al. \cite{tam91}), which drives an UC\hii 
~region (Wang et al. \cite{wan11}).\\
\indent 6.7-GHz \meth ~masers and 22-GHz \water ~masers are detected towards SMA\,1 (Goddi et al. \cite{god07}, Wang et al.
\cite{wan11}). Because the \water ~masers are associated with the inner part of the jet/outflow system ($\rm{PA_{\rm{jet}}}=67$\d, Howard et al.
 \cite{how97}; $\rm{PA_{\rm{out}}^{CO}=75}$\d, Wang et al \cite{wan11}), NIR\,3 is thought to be the driving source of this jet/outflow
 system and should not be
 older than 10$^5$~yr (Wang et al. \cite{wan11}). The velocities of the blue-shifted and red-shifted lobes of the CO-outflow are 
$-40$~\kms$<V_{\rm{S255-IR}}^{\rm{blue}}<0$~\kms ~and $+16$~\kms$<V_{\rm{S255-IR}}^{\rm{red}}<+56$~\kms, respectively (Wang et al. \cite{wan11}).
Wang et al. (\cite{wan11}) also detected a rotating toroid perpendicular to the outflow that fragmented into two sources SMA\,1 and SMA\,2.\\
\indent A Zeeman-splitting of $\Delta V_{\rm{Z}}=0.47\pm0.08$~\ms ~for the 6.7-GHz \meth ~masers was measured by Vlemmings et al. (\cite{vle11}). 
\begin{figure*}[th!]
\centering
\includegraphics[width = 8 cm]{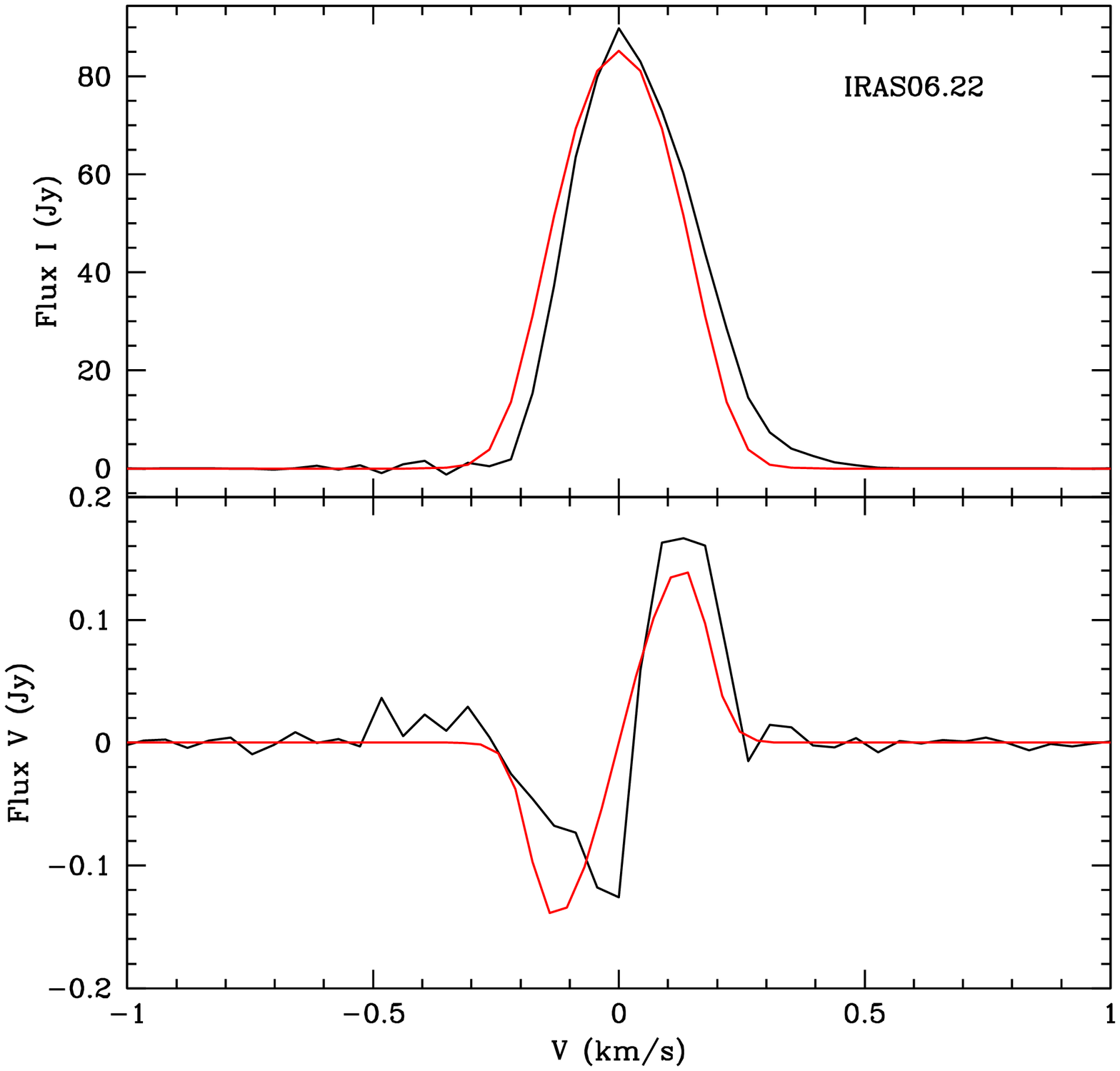}
\includegraphics[width = 8 cm]{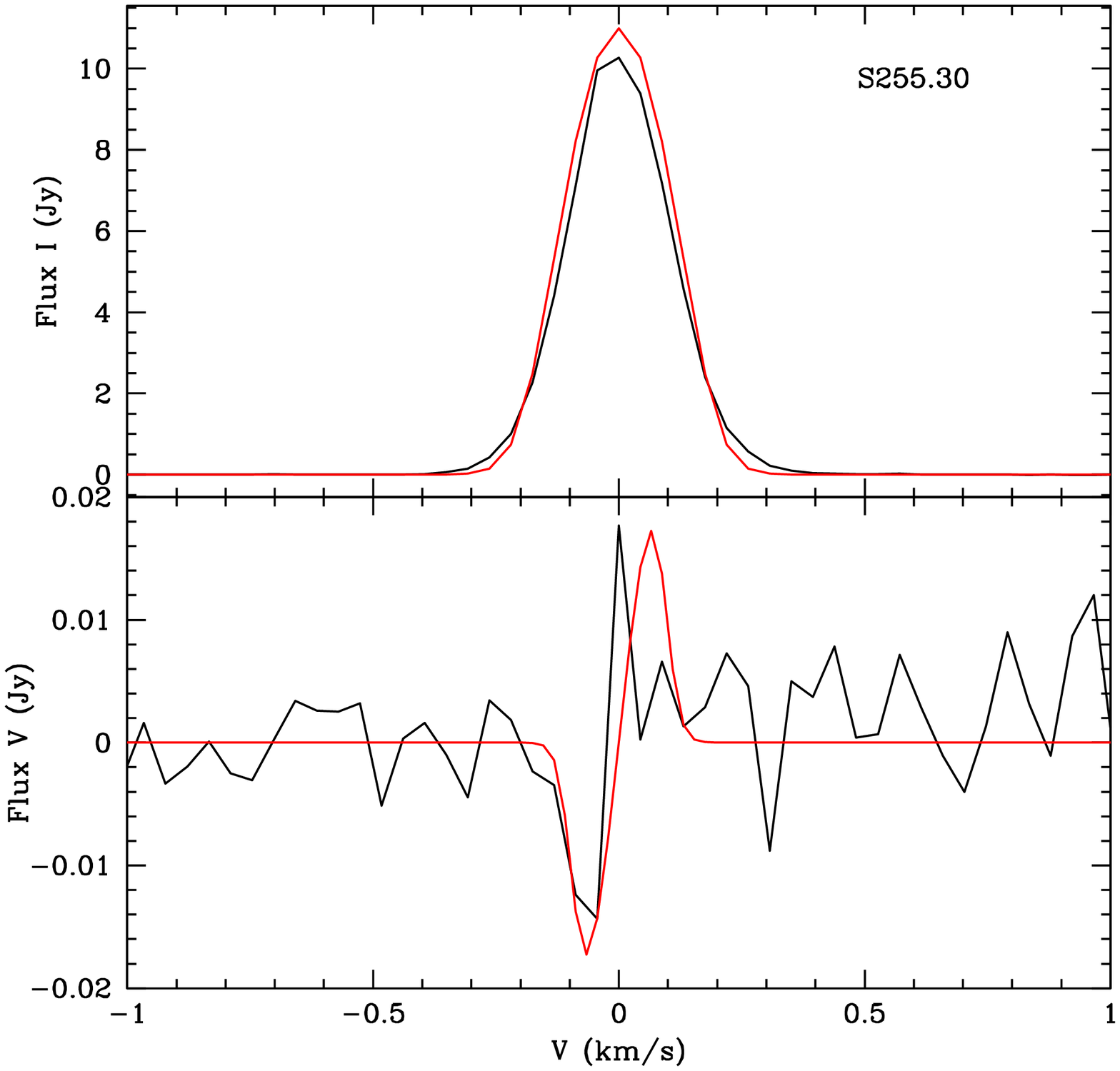}
\caption{Total intensity (\textit{I}, top) and circular polarized (\textit{V}, bottom) spectra for the two maser features 
IRAS06.22 (left panel) and S255.30  (right panel), see Tables~\ref{I06_tab} and \ref{S255_tab}. The thick red
lines are the best-fit models of \textit{I} and \textit{V} emission obtained using the adapted FRTM code (see 
Sect.~\ref{analysis}). The maser features were centered to zero velocity.}
\label{Vfit}
\end{figure*}
\subsection{S231}
IRAS05358+3543 is a cluster of embedded infrared sources associated with a number of \hii ~regions ($V_{\rm{lsr,~CO}}^{\rm{IRAS05358}}=-17.5$~\kms
, Ginsburg et al. \cite{gin09}),
 among which S231, S233, and S235 (Israel \& Felli \cite{isr78}). Even if the 6.7-GHz \meth ~maser site is surrounded by the three \hii ~regions and
 it is not directly associated with any of them, the maser site is known in literature with the name S231 (e.g. Minier et al. \cite{min00}). Following 
Heyer et al. (\cite{hey96}), we adopt a kinematic distance of 1.8~kpc for the overall complex.\\
\indent The \meth ~masers show a linear distribution that was suggested to trace an edge-on disk ($\rm{PA_{CH_{3}OH}}\approx25$\d, ~Minier et 
al. \cite{min00}), which is 
associated with one of the millimeter continuum sources identified by Beuther et al. (\cite{beu07}), i.e. mm1a. Beuther et al. (\cite{beu07}) 
suggested that mm1a, which is also associated with an hypercompact \hii ~region and a mid-infrared source, forms a binary system with mm1b 
(with a projected separation of 1700~AU). Ginsburg et al. (\cite{gin09}) observed mm1a with the VLA and they determined a projected separation 
between mm1a and the \meth ~maser site of  $\sim400$~AU suggesting that probably the binary system is formed by mm1a and the massive YSO 
associated with the \meth ~maser site. Furthermore, Ginsburg et al. (\cite{gin09}) associated one of the seven $\rm{H_2}$-outflows detected towards 
this region, the collimated outflow~2 of which only the blue-shifted lobe is visible ($\rm{PA_{out}}=133^{\circ}\pm5$\d ~,$V_{\rm{S231}}^
{\rm{blue}}\approx-47$~\kms), with the linear 
distribution of the \meth ~masers. Vlemmings et al. (\cite{vle11}) measured a Zeeman-splitting of $0.95\pm0.11$~\ms ~of the \meth ~masers.
\begin {table*}[t!]
\caption []{Observational details.} 
\begin{center}
\scriptsize
\begin{tabular}{ l c c c c c c}
\hline
\hline
\,\,\,\,\,(1)        &(2)               & (3)                           &  (4)                          &  (5)         & (6)              & (7)\\
Source               & observation      & pointing RA\tablefootmark{a}  & pointing DEC\tablefootmark{a} & calibrator   & beam size        &  rms\\ 
                     & date             & (J2000)                       & (J2000)                       &              &                  &  \\ 
                     &                  & ($\rm{^{h}:~^{m}:~^{s}}$)     & ($\rm{^{\circ}:\,':\,''}$)    &              &(mas~$\times$~mas)& (\mjyb)  \\ 
\hline
IRAS06058+2138-NIRS1 & May 30, 2011     & 06:08:53.344\tablefootmark{b} & 21:38:29.158\tablefootmark{b} & J0927+3902   & $7.3\times3.4$   & 3        \\
IRAS22272+6358A      & October 27, 2011 & 22:28:51.407\tablefootmark{b} & 64:13:41.314\tablefootmark{b} & J2202+4216   & $6.7\times4.3$   & 4         \\
S255-IR              & October 30, 2011 & 06:12:54.020\tablefootmark{b} & 17:59:23.316\tablefootmark{b} & J0359+5057   & $8.4\times3.4$   & 3         \\
S231                 & October 27, 2011 & 05:39:13.059\tablefootmark{c} & 35:45:51.29\tablefootmark{c}  & J0359+5057   & $6.5\times3.4$   & 3         \\
\hline
\end{tabular}
\end{center}
\tablefoot{
\tablefoottext{a}{The pointing position corresponds to the absolute position of the brightest maser spot measured from previous VLBI observations.}
\tablefoottext{b}{Position from Rygl et al. \cite{ryg10}.}
\tablefoottext{c}{Position from Minier et al. \cite{min00}.}
}
\label{Obs}
\end{table*}
\section{Observations and data reduction}
\label{obssect}
The sources were observed at 6.7-GHz in full polarization spectral mode with seven of the European VLBI Network\footnote{The European VLBI 
Network is a joint facility of European, Chinese, South African and other radio astronomy institutes funded by their national research 
councils.} (EVN) antennas (Effelsberg, Jodrell, Onsala, Medicina, Torun, Westerbork, and Yebes-40\,m), for a total observation time of 26~h (program code ES066). 
The bandwidth was 2~MHz, providing a velocity range of $\sim100$~\kms. The data were correlated with the EVN software correlator (SFXC) 
at Joint Institute for VLBI in Europe (JIVE) using 2048 channels and generating all 4 polarization combinations (RR, LL, RL, LR) with 
a spectral resolution of $\sim$1~kHz ($\sim$0.05~\kms). The observational details are reported in Table~\ref{Obs}.\\
\indent Because the observations
 were not performed in phase-referencing mode, we do not have information of the absolute positions of the masers. The pointing positions of our 
observations correspond to the absolute positions of the brightest maser spot of each source as measured from previous VLBI observations (see Table~\ref{Obs}).\\
\indent The data were edited and calibrated using AIPS. The bandpass, the delay, the phase, and the polarization calibration were performed 
on the calibrators listed in Table~\ref{Obs}. Fringe-fitting and 
self-calibration were performed on the brightest maser feature of each star-forming region.  The \textit{I}, \textit{Q}, 
\textit{U}, and 
\textit{V} cubes were imaged using the AIPS task IMAGR. The \textit{Q} and \textit{U} cubes 
were combined to produce cubes of polarized intensity 
($P_{\rm{l}}=\sqrt{Q^{2}+U^{2}}$) and polarization angle ($\chi=1/2\times~atan(U/Q)$). We calibrated the linear polarization angles by comparing 
the linear polarization angles of the polarization calibrators measured by us with the angles obtained by calibrating the POLCAL observations 
made by NRAO\footnote{http://www.aoc.nrao.edu/$\sim$~smyers/calibration/}. IRAS\,06058+2138 was observed a day after a POLCAL observations run
 and the polarization angle of J0927+3902 was $-85$\d$\!\!.2$. The other three sources were observed between two POLCAL observations runs during
 which the linear polarization angles were constant, the average values are $-74^{\circ}\pm4$\d ~and $-31^{\circ}\pm1$\d ~for 
J0359+5057 and J2202+4216, respectively. We were thus able to estimate the polarization angles with a systemic error of no more than 
$\sim$~5\d. The formal errors on $\chi$ are due to thermal noise. This error is given by $\sigma_{\chi}=0.5 ~\sigma_{P}/P \times 180^{\circ}/\pi$ 
(Wardle \& Kronberg \cite{war74}), where $P$ and $\sigma_{P}$ are the polarization intensity and corresponding rms error respectively.
\begin{figure*}[th!]
\centering
\includegraphics[width = 9 cm]{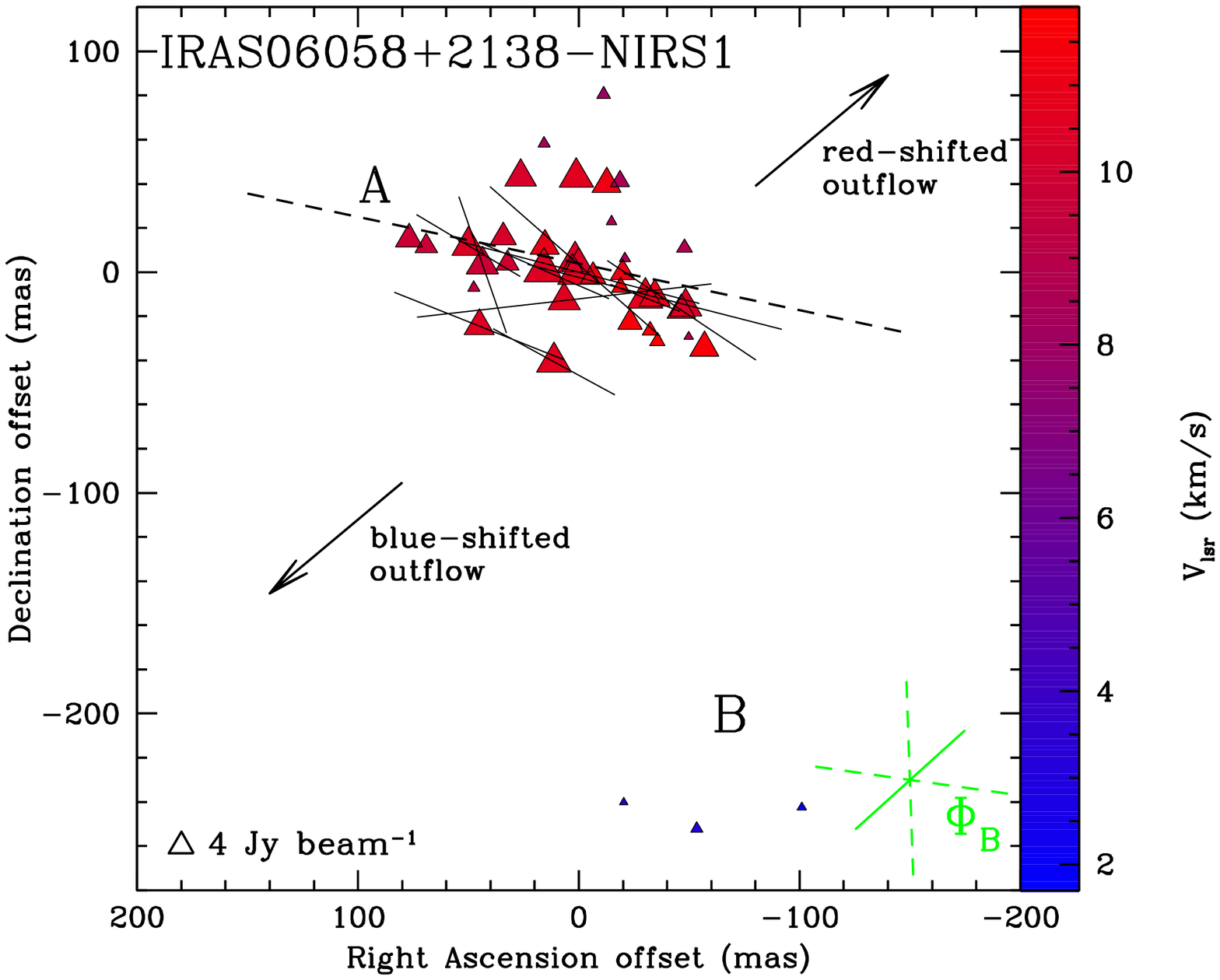}
\includegraphics[width = 9 cm]{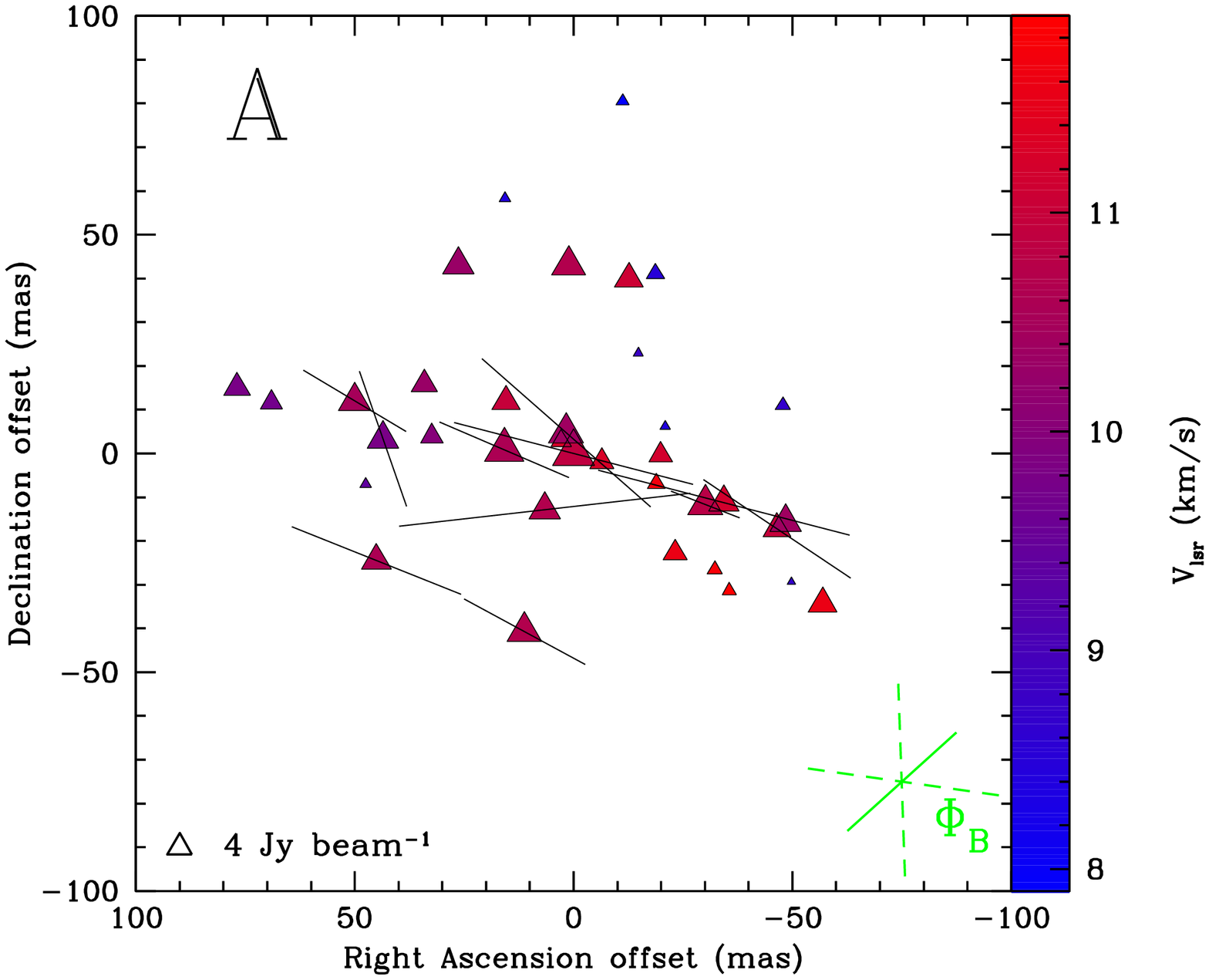}
\caption{Left panel: a view of the \meth ~maser features detected around IRAS06058+2138-NIRS\,1 Table~\ref{I06_tab}).  Triangles symbols identify \meth ~maser 
features scaled logarithmically according to their peak
flux density (Table~\ref{I06_tab}). Maser LSR radial velocities are indicated by color (the assumed velocity of the YSO is
 $V_{\rm{lsr}}^{\rm{C^{18}O}}=+3.9$~\kms, Saito et al. \cite{sai07}). A 4~\jyb ~symbol is plotted for illustration. 
The linear polarization vectors, scaled logarithmically according to polarization fraction $P_{\rm{l}}$, are overplotted. On the right-bottom 
corner the error weighted orientation of the magnetic field ($\Phi_{\rm{B}}$, see Sect.\ref{Borient}) is also reported, the two dashed segments indicates the
 uncertainty. The two arrows indicate the direction of the red- and blue-shifted lobe of the bipolar outflow ($\rm{PA_{out}}=130$\d; Wu et al.
 \cite{wu10}). The dashed line is the best linear fit of the \meth ~maser features of group~A ($\rm{PA_{CH_{3}OH}}=78^{\circ}\pm7$\d).
Right panel: a zoom-in view of group~A.}
\label{I06_cp}
\end{figure*}
\section{Analysis}
\label{analysis}
The \meth ~maser features were identified by following the same procedure described in Surcis et al. (\cite{sur11a}). 
We made use of the adapted full radiative transfer method (FRTM) code for 6.7-GHz \meth ~masers (Vlemmings et 
al. \cite{vle10}, Surcis et al. \cite{sur11b}, \cite{sur12}) to model the total intensity and the linearly polarized spectrum of every
 maser feature for which we were able to detect linearly polarized emission. \\
\indent The output of the code provides estimates of the emerging brightness temperature (\tbo), i.e. the brightness temperature that 
emerges from the maser beam, and the intrinsic thermal linewidth (\dvi), i.e. the full width half-maximum (FWHM) of the Maxwellian 
distribution of particle velocities. Note that the shapes of the total intensity, linear polarization, and circular polarization spectra of
the maser features depends on both \tbo ~and \dvi.
Following Surcis et al. (\cite{sur11b}, \cite{sur12}), we restricted our analysis to values of \dvi ~from 0.5 to 1.95~\kms.
If \tbo$>2.6\times10^9~\rm{K~sr}$, the 6.7-GHz \meth ~masers can be considered partially saturated
and their \dvi ~and \tbo ~values are, respectively, overestimated and underestimated (Surcis et al. \cite{sur11b}). However we can be confident that
the orientation of their linear polarization vectors is not affected by their saturation state (Surcis et al. \cite{sur12}), and consequently
they can be taken into account for determining the orientation of the magnetic field in the region.\\
\indent Considering \tbo ~and $P_{\rm{l}}$ we determined the angle between the maser propagation direction and the magnetic field ($\theta$). If 
$\theta>\theta_{\rm{crit}}=55$\d, where $\theta_{\rm{crit}}$ is the Van Vleck angle, the magnetic field appears to be perpendicular to the linear 
polarization vectors, otherwise it is parallel (Goldreich et al. \cite{gol73}). To better determine the orientation of the magnetic field w.r.t. the 
linear polarization vectors we take into account the errors associated with $\theta$, which we indicate here as $\varepsilon^{\rm{-}}$ and 
$\varepsilon^{\rm{+}}$, i.e. $\theta^{\varepsilon^{\rm{+}}}_{\varepsilon^{\rm{-}}}$ in Tables~\ref{I06_tab}--\ref{S231_tab}. We define the two
 limit values of the measured $\theta$ as $\theta^{\rm{-}}=\theta-\varepsilon^{\rm{-}}$ and $\theta^{\rm{+}}=\theta+\varepsilon^{\rm{+}}$. 
Considering the critical value we have $\Delta^{-}=|\theta^{\rm{-}}-55$\d$|$ and $\Delta^{+}=|\theta^{\rm{+}}-55$\d$|$. If $\Delta^{+}>\Delta^{-}$ 
the magnetic field is most likely  perpendicular to the linear polarization vectors, if $\Delta^{+}<\Delta^{-}$ the magnetic field is assumed to be 
parallel. Of course if 
$\theta^{\rm{-}}$ and $\theta^{\rm{+}}$ are both larger or smaller than 55\d ~the magnetic field is perpendicular or parallel to the linear 
polarization vectors, respectively.\\
\indent Because of technical limitations, the spectral resolution of the past observations was of about 0.1~\kms, and we were only able 
to measure the Zeeman-splitting (\dvz) 
from the cross-correlation between RR and LL spectra of the \meth ~maser features (Surcis et al. \cite{sur09, sur11b,sur12}). Nowadays,  although using
the same observing setup, the EVN SFXC at JIVE enables us to correlate the spectral data with a larger number of channels than previously possible on the
JIVE hardware correlator (Schilizzi et al. \cite{sch01}), providing a better spectral resolution (i.e., $\sim0.05$~\kms).
Because of this higher spectral resolution we can now determine the \dvz ~by using the adapted FRTM code for 6.7-GHz
\meth ~maser, as was successfully done for \water ~masers by Surcis et al. (\cite{sur11a}). The best values for \tbo ~and \dvi ~are included 
in the code to produce the \textit{I} and \textit{V} models that were used for fitting the total intensity and circular polarized spectra of the
\meth ~masers (Fig.~\ref{Vfit}). The \dvz ~measured in this way is physically more significant since the physical characteristics of the masers are
taken into account.
\section{Results}
\label{res}
In Tables~\ref{I06_tab}--\ref{S231_tab} we list all the 128 \meth ~maser features detected towards the 4 massive star-forming
regions observed with the EVN. The description of the maser distribution and the polarization results are reported for each source separately in
Sects.~\ref{I06_sec}--\ref{S231_sec}.
\subsection{\object{IRAS\,06058+2138-NIRS\,1}}
\label{I06_sec}
We list all the identified 6.7-GHz \meth ~maser features, which can be divided into two groups (A and B), in Table~\ref{I06_tab}, named as 
IRAS06.01--IRAS06.39. In the left panel of Fig.~\ref{I06_cp} we show all the 6.7-GHz \meth ~maser features while in the right panel we show only
the maser features of group~A. The three maser features of group~B were not detected in previous observations by Minier et al. (\cite{min00}). 
Group~A shows a linear distribution of about 140~mas ($\sim$250~AU) with a $\rm{PA_{CH_{3}OH}}=78^{\circ}\pm7$\d, that is almost perpendicular to the
 direction of the CO-outflow ($\rm{PA_{out}}=130$\d, Wu et al. \cite{wu10}). Although the velocity range of group~A is similar to that reported by Minier 
et al. (\cite{min00}), we do not see a clear linear velocity gradient, indicating that the masers are not likely tracing an edge-on Keplerian disk,
but they trace more complex dynamics. 
The velocities of group~A, which are red-shifted w.r.t. the velocity of NIRS\,1, fall within the velocity range of the red-shifted lobe of 
the CO-outflow and they may be related to its structure. Instead the velocities of group~B 
are slightly blue-shifted w.r.t. the velocity of NIRS\,1  but do not fall within the velocity range of the blue-shifted lobe.\\
\indent We detected linear polarization in 11 \meth ~maser features ($P_{\rm{l}}=1.3\%-9.2\%$), all of which exclusively belong to group~A.
 The adapted FRTM code was able to properly fit almost half of them. The results of the code are reported in Cols. 10, 11, and 14 of
Table~\ref{I06_tab}. Three of the maser features are partially saturated, i.e. IRAS06.09, IRAS06.22, and IRAS06.30.
 Although all the $\theta$ angles (Col. 14) are
greater than $\theta_{\rm{crit}}=55$\d, $\Delta^{+}$ is smaller than $\Delta^{-}$ for IRAS06.30 indicating that in this case the magnetic field is
 more likely parallel to the linear polarization vector. We measured a Zeeman-splitting of $\Delta V_{\rm{Z}}=3.8\pm0.6$~\ms ~for the brightest maser
 feature
 IRAS06.22, its circular polarization fraction is $P_{\rm{V}}=0.3\%$.
\subsection{\object{IRAS\,22272+6358A}}
\label{I22_sec}
\begin{figure}[t]
\centering
\includegraphics[width = 9 cm]{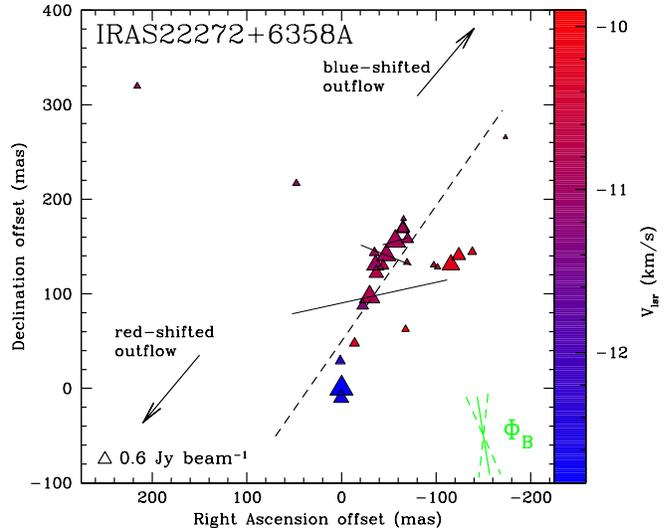}
\caption{A view of the \meth ~maser features detected around IRAS22272+6358A (Table~\ref{I22_tab}). Same symbols as in Fig.~\ref{I06_cp}.
The assumed velocity of the YSO is $V_{\rm{lsr}}=-11$~\kms ~(Beltr\'{a}n et al. \cite{bel06}).
The two arrows indicate the direction of the red- and blue-shifted lobe of the bipolar outflow ($\rm{PA_{out}}=140$\d; Beltr\'{a}n et al. \cite{bel06}).
 The dashed  line is the best linear fit of the \meth ~maser features 
($\rm{PA_{CH_{3}OH}}=145^{\circ}\pm11$\d).} 
\label{I22_cp}
\end{figure}

We detected 26 6.7-GHz \meth ~maser features towards IRAS\,22272+6358A (Fig.~\ref{I22_cp}), which are named IRAS22.01--IRAS22.26 in
Table~\ref{I22_tab}. The maser features show a linear distribution of 326~mas ($\sim$250~AU), with a position angle 
of $\rm{PA_{CH_{3}OH}}=145^{\circ}\pm11$\d, that seems
aligned with the CO-outflow ($\rm{PA_{out}}=140$\d; Beltr\'{a}n et al. \cite{bel06}). Two maser features, IRAS22.25 and IRAS22.26, are detected at a
distance of about 130~mas ($\sim$100~AU) and 330~mas ($\sim$250~AU) from the linear distribution. One of them was also detected by Rygl et al.
(\cite{ryg10}). The line-of-sight velocity of the \meth ~maser emission occurs within 2~\kms ~from the systemic velocity of the region.
\\
\indent Linearly polarized emission was detected towards 3 \meth ~maser features ($P_{\rm{l}}=0.8\%-1.7\%$), which were all successfully modeled by
the adapted FRTM code. They all appear unsaturated. The $\theta$ angles are greater than 55\d ~indicating that the magnetic field is perpendicular to
the linear polarization vectors. The rms noise of our observations did not allowed us to detect circular polarization at 5$\sigma$ for any of
the maser features ($P_{\rm{V}}<0.4\%$). 
\label{S255_sec}
\begin{figure*}[th!]
\centering
\includegraphics[width = 9 cm]{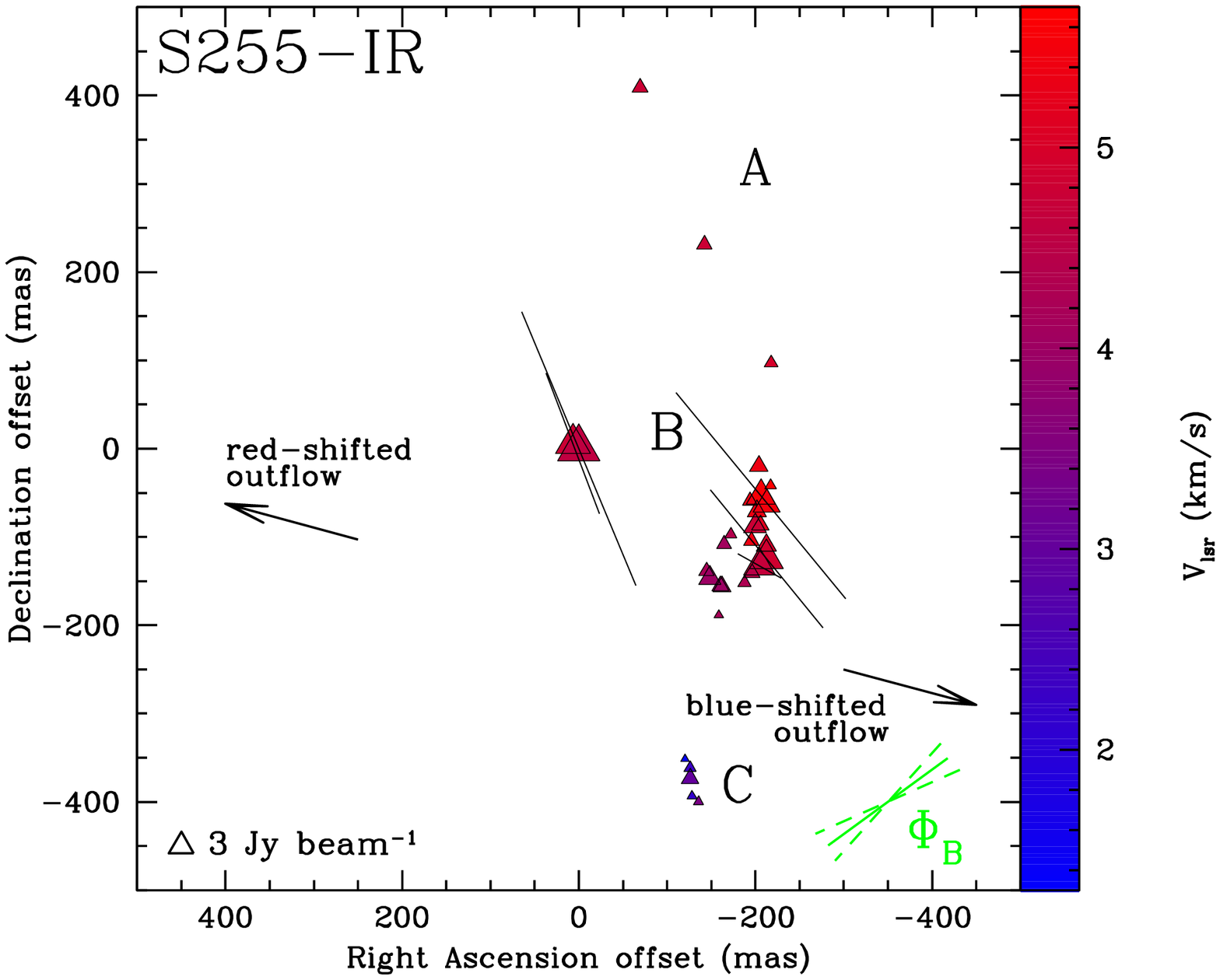}
\includegraphics[width = 9 cm]{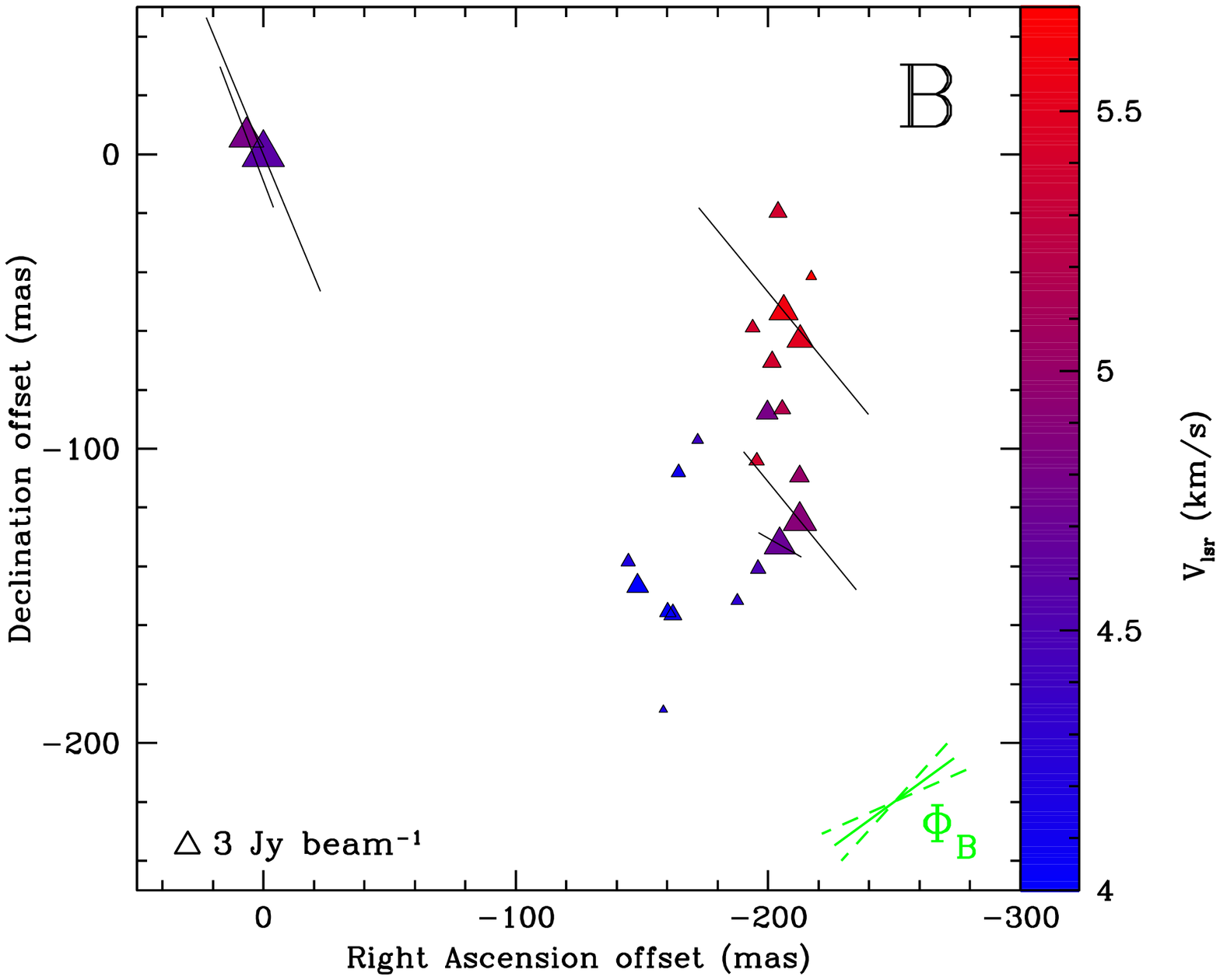}
\caption{Left panel: a view of the \meth ~maser features detected around S255-IR (Table~\ref{S255_tab}). Right panel: a zoom-in view of group~B. 
Same symbols as in Fig.~\ref{I06_cp}. The assumed velocity of the YSO is $V_{\rm{lsr}}=+5.2$~\kms ~(Wang et al. \cite{wan11}).
The two arrows indicate the direction of the red- and blue-shifted lobe of the bipolar outflow ($\rm{PA_{out}}=75$\d; Wang et al. \cite{wan11}).} 
\label{S255_cp}
\end{figure*}
\subsection{\object{S255-IR}}
In the left panel of Fig.~\ref{S255_cp} all the 31 \meth ~maser features are shown, which are listed in Table~\ref{S255_tab} (named as S255.01--
S255.31). The \meth ~maser features can be divided in three groups (A, B, and C) based on the overall spatial distribution and similar line-of-sight
velocities. While groups~B and C were also detected previously (e.g., Goddi
et al. \cite{god07}) the maser features of group~A were undetected. The overall spatial distribution of the maser emission is mainly extended on an arch
structure along the N-S direction, perpendicular to the CO-outflow ($\rm{PA_{out}}=75$\d, Wang et al. \cite{wan11}), with the most blue-shifted masers 
w.r.t. the systemic velocity of S255-IR clustered to the south.\\
\indent A linear polarization fraction between 1\% and 4.5\% was measured for 5 \meth ~maser features. The adapted FRTM code indicates that only the brightest
feature S255.30 is partially saturated because it measures \tbo$=3.2 \times 10^{9}$~K. For all the maser features $\theta>55$\d, 
so also in this case the magnetic field is perpendicular to the linear polarization vectors. A 6.7-GHz \meth ~maser Zeeman-splitting of 
\dvz$=3.2\pm0.7$~\ms ~was measured for S255.30.

\begin{figure*}[t!]
\centering
\includegraphics[width = 9 cm]{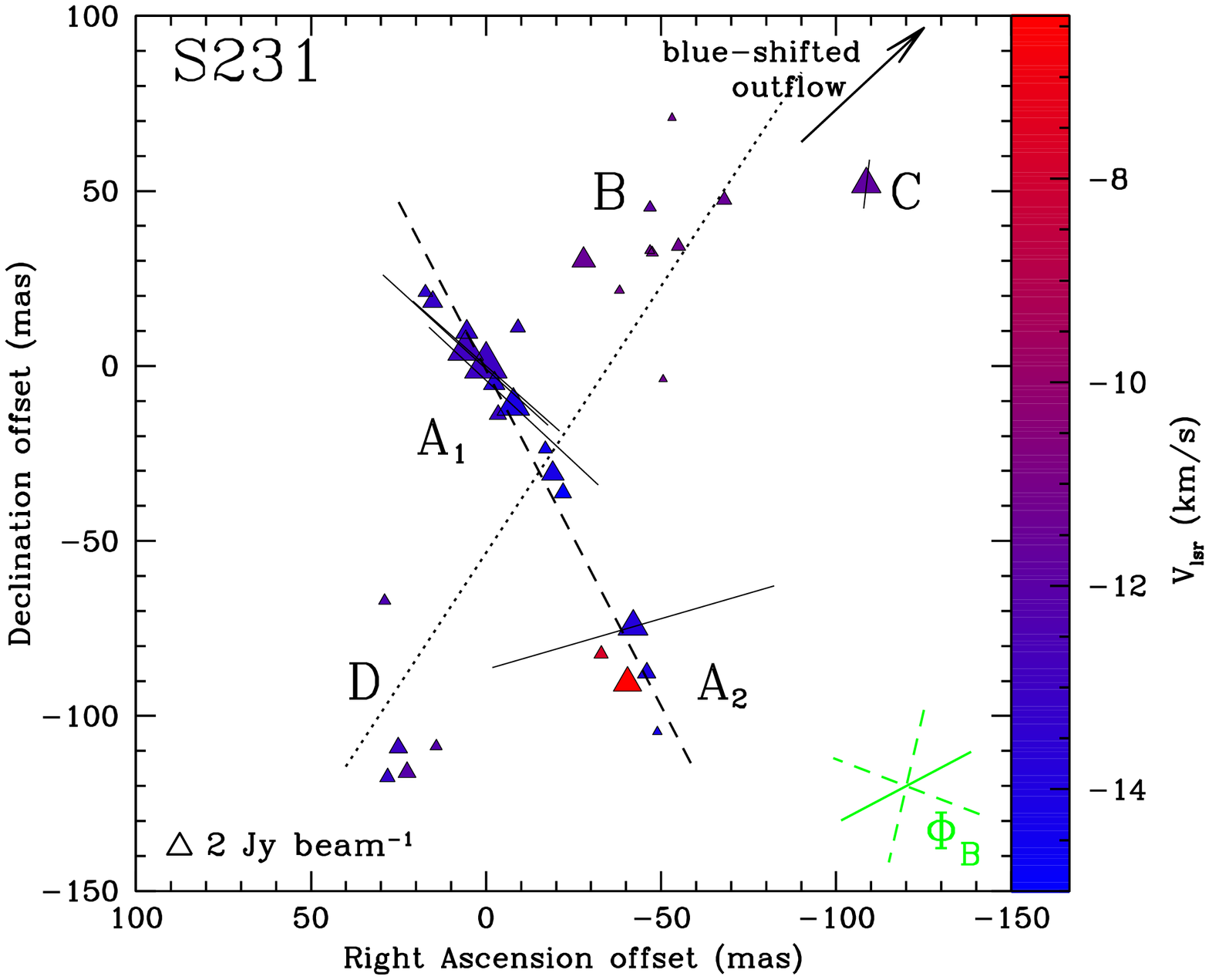}
\includegraphics[width = 9 cm]{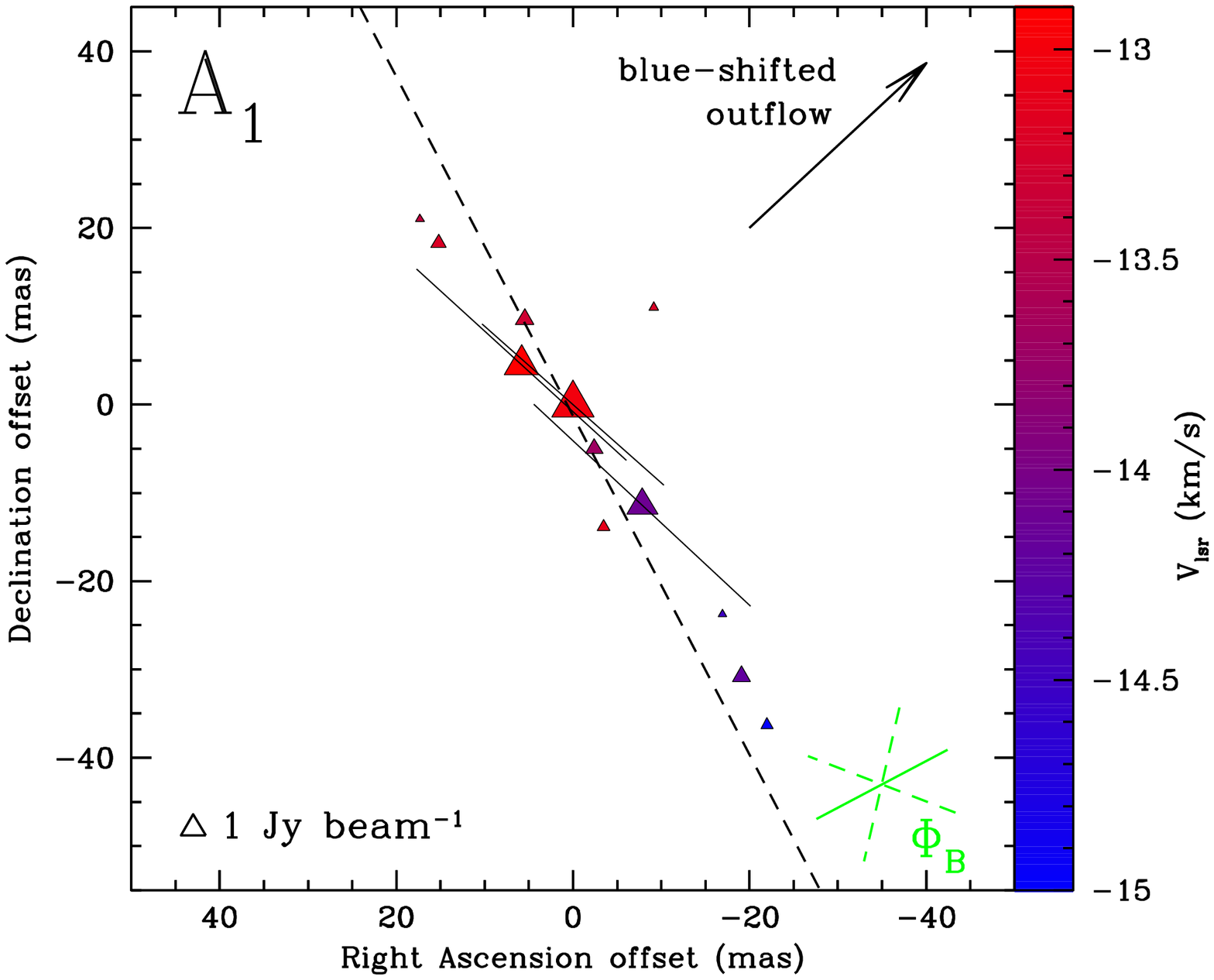}
\caption{Left panel: a view of the \meth ~maser features detected around S231 (Table~\ref{S231_tab}). Right panel: a zoom-in view of group~A1. 
Same symbols as in Fig.~\ref{I06_cp}.
 The arrow indicates the direction of the blue-shifted lobe of the $H_2$-outflow ($\rm{PA_{out}}=133$\d; Ginsburg et
 al. \cite{gin09}).The dashed  line is the best linear fit of the \meth ~maser features of group~A ($\rm{PA_{CH_{3}OH}^{A}}=28^{\circ}\pm8$\d), while the 
dotted line is  the best linear fit of the \meth ~maser features of groups~B, C, and D ($\rm{PA_{CH_{3}OH}^{BCD}}=147^{\circ}\pm12$\d) . } 
\label{S231_cp}
\end{figure*}
\subsection{\object{S231}}
\label{S231_sec}
32 \meth ~maser features were detected towards S231. They are named in Table~\ref{S231_tab} as S231.01--S231.32 and are shown in 
Fig.~\ref{S231_cp}. Following the naming convention adopted by Minier et al. (\cite{min00}), the maser features are divided in four 
groups (A, B, C, and D). Actually, group~D was not detected by Minier et al. (\cite{min00}). Furthermore, we have divided group~A in two
subgroups ($\rm{A}_{1}$ and $\rm{A}_{2}$) the second of which was also undetected by Minier et al. (\cite{min00}). The overall maser emission is 
distributed along two preferential directions. One is outlined by group~A and it is almost perpendicular to the direction of the bipolar outflow with a position
 angle of $\rm{PA_{CH_{3}OH}^{A}}=28^{\circ}\pm8$\d, which is consistent with what was measured by Minier et al. (\cite{min00}), and with a size of
 about 140~mas ($\sim250$~AU). 
A second direction is identified by combining groups~B, C, and D and has an orientation ($\rm{PA_{CH_{3}OH}^{BCD}}=147^{\circ}\pm12$\d) 
in agreement within the errors with the position angle of the outflow ($\rm{PA_{out}}=133^{\circ}\pm5$\d, Ginsburg et al. \cite{gin09}).\\
\indent Because we cannot associate the maser emission with an individual YSO unambiguously, we 
 compare the maser velocities with the average velocity of the parent molecular cloud ($V_{\rm{lsr}}^{\rm{IRAS05358}}=-17.5$~\kms). All the velocities of
 the maser features are red-shifted w.r.t. $V_{\rm{lsr}}^{\rm{IRAS05358}}$ by more than 2~\kms. \\
\indent We measured linear polarized emission from \meth ~maser features of group~A and C, with a polarization fraction 
$0.8~\%<P_{\rm{l}}<11.3~\%$, with the lowest one from group~C. From the \tbo ~values obtained using the adapted FRTM code we found that three maser features out
 of five are partially saturated, all of which are part of group~A. In S231 the magnetic field is also perpendicular to the linear polarization
vectors. No 6.7-GHz \meth ~maser Zeeman-splitting was measured in this source, unlike IRAS22272+6358A, pointing to a relatively weak 
magnetic field ($P_{\rm{V}}<0.06~\%$).
\section{Discussion}
\label{discussion}
\subsection{The linear polarization fraction and the saturation state of \meth ~masers}
\begin{figure}[th!]
\centering
\includegraphics[width = 8 cm]{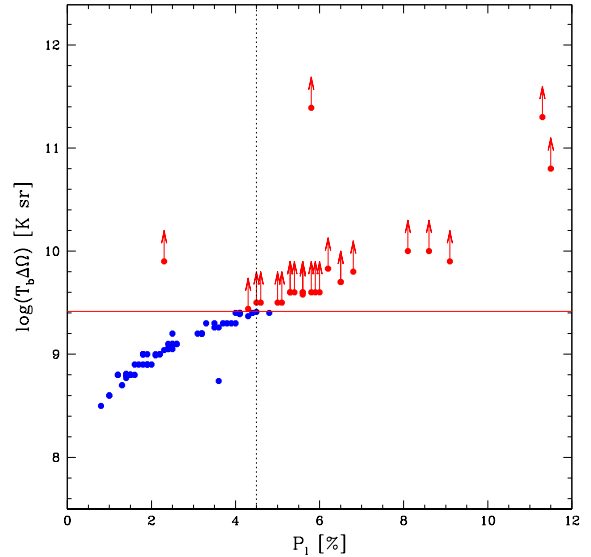}
\caption{The emerging brightness temperatures (\tbo) as function of the linear polarization fraction ($P_{\rm{l}}$). The blue 
and red circles indicate the unsaturated and saturated masers, respectively, detected in NGC7538 (Surcis et al. \cite{sur11a}), W51, W48, 
IRAS\,18556+0138, and W3(OH) (Surcis et al. \cite{sur12}), IRAS06058+2138-NIRS1, IRAS22272+6358A, S255-IR, and S231 (this work). 
The red arrows indicate that the \tbo ~values obtained from the adapted FRTM code are lower limits. The red full 
line is the limit of emerging brightness temperature above which the \meth ~masers are considered saturated 
($T_{\rm{b}}\Delta\Omega>2.6\times10^9$~K~sr; Surcis et al. \cite{sur11a}), and the dotted line gives the lower limit to the linear polarization
 fraction for saturated masers ($P_{\rm{l}}\approx4.5~\%$, Surcis et al. \cite{sur12}).}
\label{pltb}
\end{figure}

Modeling the total and the linearly polarized intensity of \meth ~masers with the adapted FRTM code enables us not only to derive the orientation
 of the magnetic fields and to measure the Zeeman-splittings accurately, but also to estimate the saturation state of the \meth ~masers.\\
\indent As we have already mentioned in Sect.~\ref{analysis}, the 6.7-GHz \meth ~maser can be considered partially saturated if 
\tbo$~>2.6\times10^9~\rm{K~sr}$. Surcis et al. (\cite{sur11b}) derived this value by considering the stimulated emission rate ($R$) and the theoretical 
condition for masers unsaturated or saturated. We refer the reader to Surcis et al. (\cite{sur11b}) for more details. The information 
of the
unsaturated or saturated state of the maser was used by Surcis et al. (\cite{sur12}) to determine a correlation between the saturation state of the
\meth ~masers and their linear polarization fraction ($P_{\rm{l}}$). They found that the 6.7-GHz \meth ~masers with $P_{\rm{l}}\lesssim4.5~\%$ are 
unsaturated. Adding the 19 new \meth ~masers that we have detected in this work, for which it has been possible to determine \tbo, to the previous 
measured 72 masers we confirm the value of 4.5~\% as found by Surcis et al. (\cite{sur12}) but with the exception of IRAS06.30 (see Fig.~\ref{pltb} and 
Table~\ref{I06_cp}). The model of IRAS06.30 is probably influenced by the brightest maser (IRAS06.22), which is located within 15~mas. 
This maser is twice a bright and highly linearly polarized.\\
\subsection{Magnetic fields in the second EVN group}
\begin {table*}[th!]
\caption []{Comparison between position angle of magnetic field, \meth ~maser distribution, outflows, and linear polarization angles.} 
\begin{center}
\scriptsize
\begin{tabular}{ l c c c c c c c c c}
\hline
\hline
\,\,\,\,\,(1)&(2)            & (3)                           & (4)                  & (5)                       & (6)        &                  (7)                              & (8)                                          & (9)                                            &(10)\\
Source &$\langle\chi\rangle$ & $\langle\Phi_{\rm{B}}\rangle$ & $\rm{PA}_{\rm{out}}$ & $\rm{PA}_{\rm{CH_{3}OH}}$ & $\rho$     & $|\rm{PA}_{\rm{out}}-\langle\Phi_{\rm{B}}\rangle|$& $|\rm{PA}_{\rm{CH_{3}OH}}-\langle\chi\rangle|$& $|\rm{PA}_{\rm{CH_{3}OH}}-\rm{PA}_{\rm{out}}|$&ref.\tablefootmark{a} \\ 
       & (\d)                & (\d)                          & (\d)                 & (\d)                      &            & (\d)                                              &(\d)                                          & (\d)                                  & \\ 
\hline
Cepheus~A & $-57\pm28$       & $+30\pm19$\tablefootmark{b}   & $+40\pm4$\tablefootmark{c} & $-79\pm9$           &  -0.34     &$10\pm19$                                          & $22\pm29$                                    & $61\pm10$\tablefootmark{d}       & (1), (2)\\
W75N-group~A & $-13\pm9$     & $+77\pm9$\tablefootmark{b}    & $+66\pm15$\tablefootmark{e} & $+43\pm10$         &  +0.96     & $11\pm18$                                         & $56\pm14$                                    & $23\pm18$                        & (3), (4)\\
NGC7538-IRS1 & $-30\pm69$    & $+67\pm70$\tablefootmark{b}   & $-40\pm10$\tablefootmark{f} & $+84\pm7$\tablefootmark{g} & +0.15& $73\pm71$\tablefootmark{d}                      & $66\pm69$\tablefootmark{d}                   & $56\pm12$\tablefootmark{d}       & (5), (6), (7)\\
W3(OH)-group II & $+21\pm45$ & $-47\pm44$\tablefootmark{b}   & $-$                  & $-59\pm6$                 &  -0.84     & $-$                                               & $80\pm45$                                    & $-$                              & (8)\\
W51-e2    & $+33\pm16$       & $-60\pm21$\tablefootmark{b}   & $-50\pm20$\tablefootmark{h} & $+57\pm8$          &  +0.70     & $10\pm29$                                         & $24\pm18$                                    & $73\pm22$\tablefootmark{d}       & (8), (9), (10)\\
IRAS18556+0138 & $-2\pm11$   & $+88\pm11$\tablefootmark{b}   & $+58\pm23$\tablefootmark{i}  & $-40\pm2$         &  -0.99     & $30\pm26$                                         & $42\pm11$                                    & $82\pm23$\tablefootmark{d}       & (8), (11)\\
W48       & $+23\pm7$        & $-67\pm7$\tablefootmark{b}    & $-$                  & $+55\pm10$                &  +0.70     & $-$                                               & $78\pm12$                                    & $-$                              & (8) \\
IRAS06058+2138-NIRS1 & $+49\pm47$ & $-49\pm52$\tablefootmark{b} & $-50\pm15$\tablefootmark{j}  & $+78\pm7$      &  +0.64     & $1\pm54$                                          & $29\pm48$                                    & $52\pm17$\tablefootmark{d}       & (12), (13)\\
IRAS22272+6358A & $-80\pm15$ & $+9\pm15$\tablefootmark{b}    & $-40\pm15$\tablefootmark{e} & $-35\pm11$         &  -0.87     & $49\pm21$                                         & $45\pm19$                                    & $5\pm19$                         & (12), (14) \\
S255-IR   & $+36\pm12$       & $-54\pm12$\tablefootmark{b}   & $+75\pm15$\tablefootmark{e} & $-63\pm49$\tablefootmark{k}& -0.11 &$51\pm19$\tablefootmark{d}                      & $81\pm51$\tablefootmark{d}                   & $42\pm51$\tablefootmark{d}       & (12), (15) \\
S231      & $+28\pm49$       & $-62\pm49$\tablefootmark{b}   & $-47\pm5$            & $+28\pm8$                 &  +0.97     & $15\pm49$                                         & $0\pm50$                                     & $75\pm9$                         & (12), (16)\\
G291.27-0.70 & $-32\pm5$     & $+52\pm5$                     & $-$                  & $-77\pm14$\tablefootmark{l}&  $-$      & $-$                                               & $45\pm15$                                    & $-$                              & (17)\\
G305.21+0.21 & $-51\pm14$    & $28\pm14$                     & $-$                  & $+48\pm23$\tablefootmark{m}&  $-$      &$-$                                                & $81\pm27$\tablefootmark{d}                   & $-$                              & (17), (18)\\
G309.92+0.47 & $+2\pm56$     & $-75\pm56$                    & $-$                  & $+35\pm5$\tablefootmark{m}&   $-$      &$-$                                                & $33\pm56$                                    & $-$                              & (17), (18)\\
G316.64-0.08 & $-67\pm36$    & $+21\pm36$                    & $-$                  & $+34\pm29$\tablefootmark{l}&  $-$      & $-$                                               & $79\pm46$\tablefootmark{d}                   & $-$                              & (17)\\
G335.79+0.17 & $+44\pm28$    & $-41\pm28$                    & $-$                  & $-69\pm25$\tablefootmark{m}&  $-$      & $-$                                               & $67\pm38$\tablefootmark{d}                   & $-$                              & (17), (18) \\
G339.88-1.26 & $+77\pm24$    & $-12\pm24$                    & $-$                  & $-60\pm17$\tablefootmark{m}&  $-$      & $-$                                               & $43\pm29$\tablefootmark{d}                   & $-$                              & (17), (18) \\
G345.01+1.79 & $+5\pm39$     & $-86\pm39$                    & $-$                  & $+74\pm4$\tablefootmark{m}&   $-$      &$-$                                                & $69\pm39$                                    & $-$                              & (17), (18) \\
NGC6334F (central) & $+77\pm20$ & $-13\pm20$                 & $-$                  & $-41\pm16$\tablefootmark{l}&  $-$      & $-$                                               & $62\pm26$\tablefootmark{d}                   & $-$                              & (17)\\
NGC6334F (NW)& $-71\pm20$    & $+19\pm20$                    & $-$                  & $-80\pm38$\tablefootmark{l}&  $-$      &$-$                                                & $9\pm43$                                     & $-$                              & (17)\\
\hline
\end{tabular}
\end{center}
\tablefoot{
\tablefoottext{a}{(1) Vlemmings et al. (\cite{vle10}); Curiel et al. (\cite{cur06}); (3) Surcis et al. (\cite{sur09}); (4) Hunter et al. (\cite{hun94});
(5) Surcis et al. (\cite{sur11b});(6) Scoville et al. (\cite{sco86}); (7) Kameya et al. (\cite{kam89}); (8) Surcis et al. (\cite{sur12}); 
(9) Keto \& Klaassen (\cite{ket08}); (10) Zhang et al. (\cite{zha98}); (11) Gibb et al. (\cite{gib03}); (12) This work; (13) Wu et al. (\cite{wu10});
(14) Beltr\'{a}n et al. (\cite{bel06}); (15) Wang et al. (\cite{wan11}); (16) Ginsburg et al. (\cite{gin09}); (17) Dodson \& Moriarty (\cite{dod12}); 
(18) De Buizer (\cite{deb03})}.
\tablefoottext{b}{Before averaging we use the criterion described in Sect.~\ref{analysis} to estimate the orientation of the magnetic field w.r.t the 
linear polarization vectors.}
\tablefoottext{c}{It has been obtained by flux weighting the angles reported in Table~2 of Curiel et al. (\cite{cur06}).}
\tablefoottext{d}{The differences between the angles are evaluated taking into account that $\rm{PA}\equiv\rm{PA}\pm180$\d,  
$\langle\chi\rangle\equiv\langle\chi\rangle\pm180$\d, and $\langle\Phi_{\rm{B}}\rangle\equiv\langle\Phi_{\rm{B}}\rangle\pm180$\d.}
\tablefoottext{e}{We consider an arbitrary conservative error of 15\d.}
\tablefoottext{f}{The errors are evaluated considering the minimum and maximum PA of the CO-outflows reported by Scoville et al. (\cite{sco86}).}
\tablefoottext{g}{We do not consider group E of \meth ~masers in the fit.}
\tablefoottext{h}{Keto \& Klaassen (\cite{ket08}) reported that the orientation of the outflow is along the rotation axis of the molecular accretion flows.
So we evaluate the arithmetic mean value ($\rm{PA_{out}}$) of the rotation axes measured from the CH$_{3}$CN ($\rm{PA_{CH_{3}CN}}=110$\d; Zhang et al. 
\cite{zha98}) and the H53$\rm{\alpha}$ flows ($\rm{PA_{H53\alpha}}=150$\d; Keto \& Klaassen \cite{ket08}).}
\tablefoottext{i}{We overestimate the errors by considering half of the opening angle of the outflow.}
\tablefoottext{j}{The $\rm{PA_{out}}$ has been estimated from Fig.~4 of Wu et al. (\cite{wu10}).}
\tablefoottext{k}{S255.30 and S255.31 were not included in the fit.}
\tablefoottext{l}{The errors are overestimated by considering Figs. 1, 5, 9, and 10 of Dodson \& Moriarty (\cite{dod12}).}
\tablefoottext{m}{The errors are equal to the difference between the angles determined by Dodson \& Moriarty (\cite{dod12}) and by De Buizer (\cite{deb03}).}
}
\label{Comp_ang}
\end{table*}
\subsubsection{Magnetic field orientations}
\label{Borient}
Before discussing the orientation of magnetic fields in the four massive YSOs, it is important to estimate whether the medium between the source
and the observer could produce a significant rotation of the linear polarization vectors. This phenomenon is known as foreground Faraday rotation 
($\Phi_{\rm{f}}$) and, as derived for the 6.7-GHz \meth ~maser by Surcis et al. (\cite{sur12}), it can be written as
\begin{equation}
\Phi_{\rm{f}}[^{\circ}]=2.26~\left(\frac{D}{[\rm{kpc}]}\right),
\label{eq10}
\end{equation}
where they assumed for the homogeneous interstellar electron density $n_{\rm{e}}\approx0.012\,\rm{~cm^{-3}}$ and for the interstellar magnetic fields
$B_{||}\approx2\,\rm{~\mu G}$. Because the massive star-forming regions investigated so far are at a distance of a few kpc, $\Phi_{\rm{f}}$
 is estimated to be within the errors of the linear polarization angles of the \meth ~maser emission (Surcis et al. \cite{sur09, sur11b, sur12}). 
This is also true in the case of the four YSOs presented in this work, for which $\Phi_{\rm{f}}$ is estimated to range between about 2\d ~and 4\d.\\
\indent Another effect that may affect the measurements of the linear polarization vectors is the internal Faraday rotation. Surcis et al. 
(\cite{sur12}) argued
 that because the linear polarization vectors of 6.7-GHz \meth ~masers are quite accurately aligned in each source, the internal 
Faraday rotation should be negligible. This general result applies also to the sources presented here. Furthermore, the saturation
 state of the masers seems to not affect the linearly polarized emission of \meth ~masers as also noted by Surcis et al. (\cite{sur12}).\\
\indent In the following, we discuss separately the orientation of the magnetic field in each of the current sources.\\

\noindent \textit{\textbf{IRAS06058+2138-NIRS1.}} We measured $\theta$ angles for six out of 11 \meth ~masers. For the \meth ~masers for which we could 
not measure the $\theta$ angles we supposed that $\theta>55$\d, as generally found (Surcis et al. \cite{sur11b, sur12}). Moreover, we measured 
$\Delta^{+}<\Delta^{-}$ only for the \meth ~maser IRAS06.30, which means that the magnetic field is parallel to its linear polarization vector (see 
Sect.~\ref{analysis}). As a result, the error weighted orientation of the magnetic field is $\langle\Phi_{\rm{B}}\rangle=131^{\circ}\pm52$\d 
~indicating that the magnetic field is along the bipolar CO-outflow ($\rm{PA_{out}}=130$\d; Wu et al. \cite{wu10}). Remembering that if 
$\Delta V_{\rm{Z}}>0$ the magnetic field is pointing away from the observer and if $\Delta V_{\rm{Z}}<0$ towards the observer, from the 
Zeeman-splitting measurements we can also state that the magnetic field is pointing away from us.\\

\noindent \textit{\textbf{IRAS22272+6358A.}} The magnetic field is perpendicular to all the linear polarization vectors measured towards the source,
i.e. $\langle\Phi_{\rm{B}}\rangle=9^{\circ}\pm15$\d. The orientation of the magnetic field is rotated by about 50\d, i.e. within $3~\sigma$
, w.r.t. the direction of both the CO-outflow and the linear distribution of the \meth ~masers. The magnetic field might be along  
the extended north-south component of the C$^{18}$O emission detected by Beltr\'{a}n et al. (\cite{bel06}). By looking at each single linear 
polarization vectors (Fig.~\ref{I22_cp}), we speculate that the \meth ~masers might probe a twisted magnetic 
field around the linear distribution of masers, or around the CO-outflow.\\

\noindent \textit{\textbf{S255-IR.}} According to the linear polarization vectors and to the measured $\theta$ angles, the magnetic field has an error 
weighted orientation of  $\langle\Phi_{\rm{B}}\rangle=126$\d~$\pm~12$\d. In this case the magnetic field is not aligned with the CO-outflow
 ($\rm{PA_{out}}=75$\d; 
Wang et al. \cite{wan11}) but it may be associated with the rotating torus ($\rm{PA_{torus}}=165$\d; Wang et al. \cite{wan11}), like 
what was measured in NGC7538-IRS1 (Surcis et al. \cite{sur11b}). Also in this case, according to \dvz, 
the magnetic field is pointing away from us. Since the spatial distribution of the \water ~masers in S255-IR is well aligned with the outflow
 direction (Goddi et al. \cite{god07}), interferometric polarization observations of \water ~masers will be crucial for interpreting the 
overall magnetic field morphology.\\

\noindent \textit{\textbf{S231.}} The error-weighted orientation of the magnetic field is $\langle\Phi_{\rm{B}}\rangle=118^{\circ}\pm49$\d. Despite 
that the angle 
of the linear polarization vector of S231.11 is more than one hundred degrees misaligned with the other vectors, the magnetic field is
 along both to the CO-outflow ($\rm{PA_{out}}=133^{\circ}\pm5$\d; Ginsburg et al. \cite{gin09}) and to the linear fit of groups~B, C, and D 
($\rm{PA_{CH_{3}OH}^{BCD}}=147^{\circ}\pm12$\d; see Sect.~\ref{S231_sec}).
\begin{figure*}[th!]
\centering
\includegraphics[width = 8 cm]{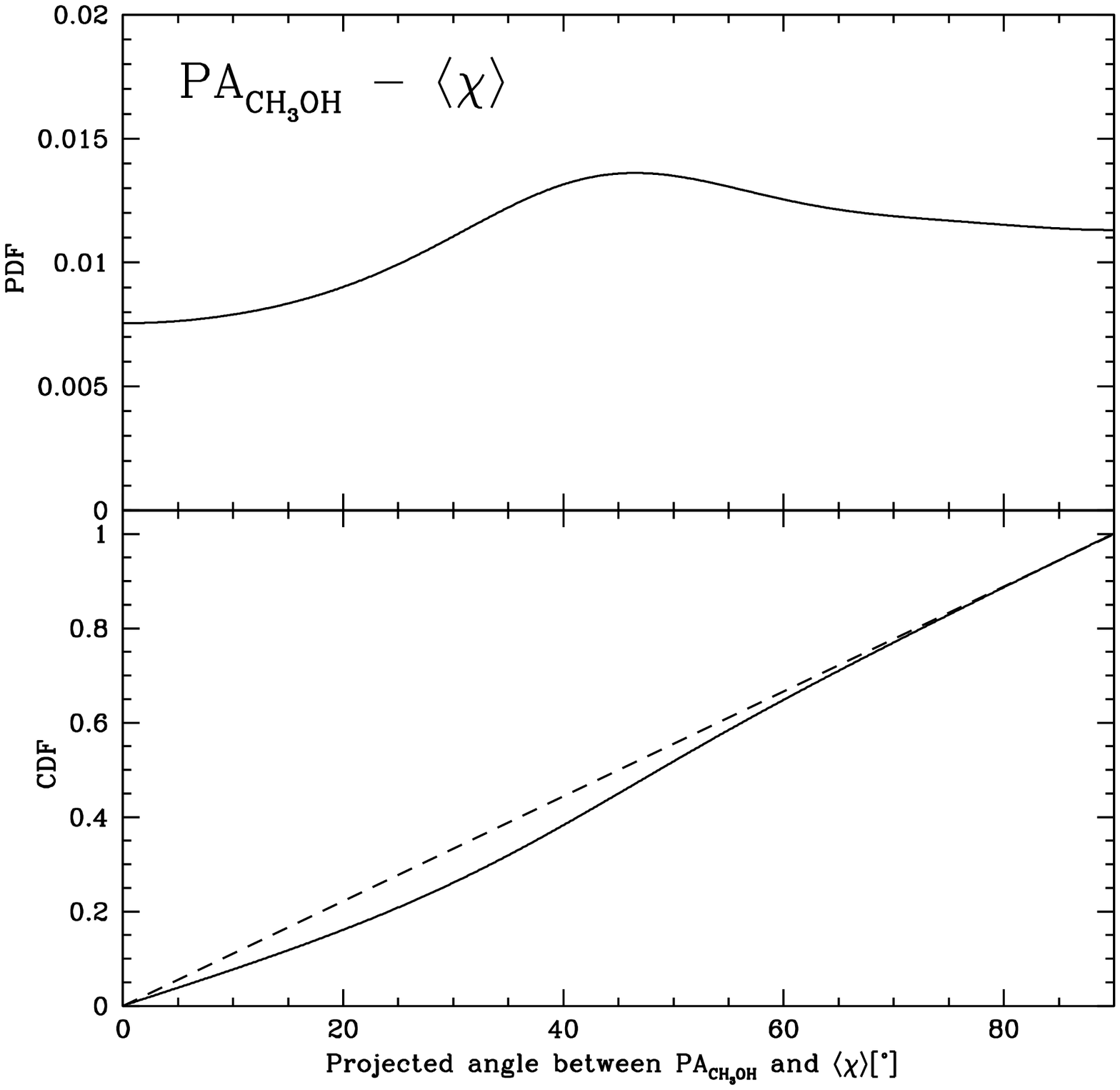}
\includegraphics[width = 8 cm]{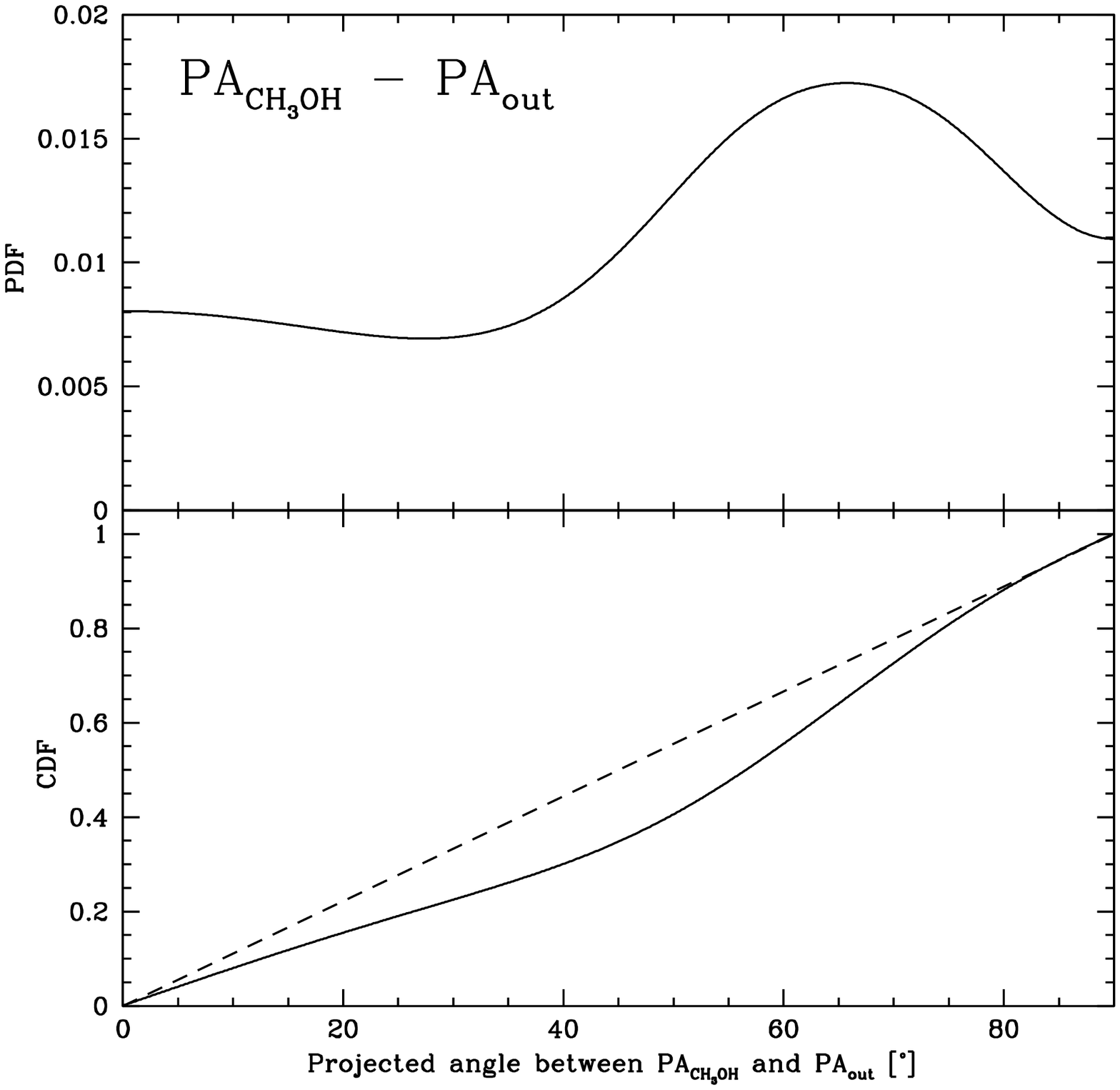}
\caption{\footnotesize{Left:The probability distribution function (PDF, top panel) and the cumulative distribution function (CDF, bottom panel) of the
projected angle between the PA of the \meth ~maser distribution and the linear polarization angles 
($|\rm{PA}_{\rm{CH_{3}OH}}-\langle\chi\rangle|$). Right: The PDF and the CDF of the projected angle between the PA of the \meth ~maser
 distribution and the outflow axes ($|\rm{PA}_{\rm{CH_{3}OH}}-\rm{PA_{\rm{out}}}|$). In both panels the dashed line is the 
CDF for random orientation of outflows and magnetic fields, i.e. all angular differences are equally likely.The results of the 
Kolmogorov-Smirnov test can be found in Table~\ref{KS}}}
\label{cdf}
\end{figure*}
\subsection{The importance of magnetic fields in high-mass star formation.}
The importance of magnetic field in star-forming regions can be estimated by evaluating the ratio $\beta$ between thermal and magnetic energies.
If $\beta<1$ the magnetic field is dynamically important. To measure $\beta$ we need to know the magnetic field strength. As we mentioned
in Sect.~\ref{intro} the Land\'{e} g-factor of the 6.7-GHz \meth ~transition, from which $\alpha_{\rm{Z}}$ is evaluated, is still 
uncertain (Vlemmings et al. \cite{vle11}). Consequently we can only parametrize the ratio $\beta$ as function of $\alpha_{\rm{Z}}$ by considering
the eq.~2 of Surcis et al. (\cite{sur12})
\begin{equation}
 \beta=611.6~\alpha_Z^{2} ~cos~\langle\theta\rangle ~\left(\frac{|\Delta V_{\rm{Z}}|}{[\rm{ms^{-1}}]}\right)^{-2}.
\label{eq5}
\end{equation}
We find that $\beta_{\rm{IRAS06.22}}\simeq12.4\times\alpha_Z^{2}$ and $\beta_{\rm{S255.30}}\simeq8\times\alpha_Z^{2}$. Considering a 
 reasonable value for $\alpha_{\rm{Z}}$ in the range $0.005$~\kmsg~$<\alpha_{\rm{Z}}<0.05$~\kmsg ~(Surcis et al.
 \cite{sur11b}), we have a range of $\beta$ between $10^{-4}$ and $10^{-2}$ for both sources. 
As it was found for all the previous massive star-forming regions
investigated using the 6.7-GHz \meth ~maser emission (Surcis et al. \cite{sur09, sur11b, sur12}), magnetic fields play an important role in the 
dynamics of IRAS06058+2138-NIRS1 and S255-IR.
\subsection{Magnetic fields and outflows.}
\label{MFO}
In the last four years a total of 20 massive star-forming regions were observed at mas resolution at 6.7-GHz all over the sky to measure the 
orientation of magnetic fields around massive YSOs. They all are listed in Table~\ref{Comp_ang}. In the northern hemisphere the sources were
observed using the
\begin{figure}[t!]
\centering
\includegraphics[width = 8 cm]{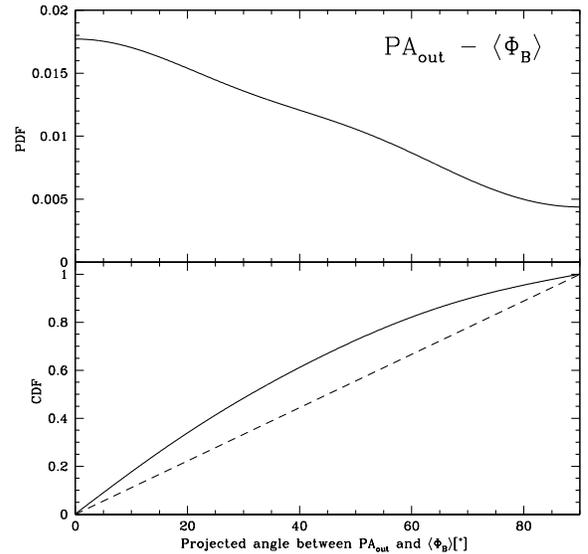}
\caption{\footnotesize{ The probability distribution function (PDF, top panel) and the cumulative distribution function (CDF, bottom panel) of the
projected angle between the magnetic field and the outflow axes ($|\rm{PA}_{\rm{out}}-\langle\Phi_{\rm{B}}\rangle|$). The dashed line is the 
CDF for random orientation of outflows and magnetic fields, i.e. all angular differences are equally likely. The results of the 
Kolmogorov-Smirnov test can be found in Table~\ref{KS}}}
\label{cdf2}
\end{figure}
EVN (Surcis et al. \cite{sur09, sur11b,sur12}, and this work) and the Multi-Element Radio Linked Interferometer network (MERLIN) (Vlemmings et
al. \cite{vle10}), while in the southern hemisphere by using the Australia Telescope Compact Array (ATCA) (Dodson \& Moriarty \cite{dod12}). 
This total sample, which we refer to as the magnetic field total sample, gives us the opportunity to start a statistical analysis of the 
results so far.\\
\indent We compare the orientation of the \meth ~maser distribution ($\rm{PA}_{\rm{CH_{3}OH}}$, Col.~5 of Table~\ref{Comp_ang}) with the 
error weighted value of the linear polarization angles ($\langle\chi\rangle$, Col.~2) and with the orientation of the molecular outflows 
($\rm{PA}_{\rm{out}}$, Col.~4), i.e. $\mathbf{|\rm{PA}_{\rm{CH_{3}OH}}-\langle\chi\rangle|}$ (Col. 8) and 
$\mathbf{|\rm{PA}_{\rm{CH_{3}OH}}-\rm{PA}_{\rm{out}}|}$ (Col.9) respectively. In Col.~6 the correlation coefficients ($\rho$) of the linear
 fits made by us are also reported. For our statistical analysis we require the uncertainties of all the angles. While the errors of 
$\rm{PA}_{\rm{CH_{3}OH}}$ and $\langle\chi\rangle$ are easily determined, the uncertainties of $\rm{PA}_{\rm{out}}$ are unknown for most of the 
sources, so either, when possible, we estimate them (see footnotes of Table~\ref{Comp_ang}) or we consider a conservative uncertainty of
 $\pm15$\d. Moreover, we also compare the error weighted orientation of the magnetic field ($\langle\Phi_{\rm{B}}\rangle$, Col.~3) 
with $\rm{PA}_{\rm{out}}$, i.e $\mathbf{|\rm{PA}_{\rm{out}}-\langle\Phi_{\rm{B}}\rangle|}$ in Col.~7, for the 9 sources for which an outflow 
has been detected. Note that all the angles in Table~\ref{Comp_ang} are the projection on the plane of the sky. The uncertainties in Cols. 7,
 8, and 9 are equal to $\sigma_{\rm{x-y}}=\sqrt{\sigma^{2}_{\rm{x}}+\sigma^{2}_{\rm{y}}}$, where x and y are the two angles taken in consideration 
in each column.\\
\indent Fig.~\ref{cdf} shows the probability distribution function (PDF) and the cumulative distribution function (CDF) of the projected angles
  $|\rm{PA}_{\rm{CH_{3}OH}}-\langle\chi\rangle|$ and $|\rm{PA}_{\rm{CH_{3}OH}}-\rm{PA}_{\rm{out}}|$. Because the angles for the different
 sources have different errors, which need to be taken into account, we treat in Fig.~\ref{cdf} the full distributions instead of the histograms.
The CDFs show that there is currently no indication of a relation between $\rm{PA}_{\rm{CH_{3}OH}}$ and the angles $\langle\chi\rangle$ and 
$\rm{PA}_{\rm{out}}$. Indeed, a Kolmogorov-Smirnov (K-S) test shows that the probabilities that 
$|\rm{PA}_{\rm{CH_{3}OH}}-\langle\chi\rangle|$ and $|\rm{PA}_{\rm{CH_{3}OH}}-\rm{PA_{\rm{out}}}|$ are drawn from random distributions
(dashed lines in Fig.~\ref{cdf}), i.e. all angular differences are equally likely, are in both cases $\sim$60~\%. In Table~\ref{KS}  the results
 of the K-S test are reported, where $D$ (Col.~3) is the maximum value of the absolute difference between the data set, composed of $N$ (Col.~2) 
elements, and the random distribution, and
\begin{equation}
Q_{\rm{K-S}}(\lambda)=2\sum_{j=1}^{N} (-1)^{j-1}~ e^{-2j^2\lambda^2}
\end{equation}
in Col.~5 is the significance level of the test. Here $\lambda=(\sqrt{N}+0.12+0.11/\sqrt{N})\times D$ (Col.~4). The K-S test becomes 
asymptotically accurate as $N$ becomes large.
\begin {table}[t]
\caption []{Results of Komogorov-Smirnov test.} 
\begin{center}
\scriptsize
\begin{tabular}{ l c c c c }
\hline
\hline
\,\,\,\,\,\,\,\,\,\,\,\,\,\,\,(1)                &(2)   & (3)  & (4)       & (5)          \\ 
\,\,\,\,\,\,\,\,\,\,Angle                        & $N$  & $D$    & $\lambda$ & $Q_{\rm{K-S}}(\lambda)$\\
\hline
\\
$|\rm{PA}_{\rm{CH_{3}OH}}-\langle\chi\rangle|$   & 20   & 0.17 & 0.78      & 0.60 \\
$|\rm{PA}_{\rm{CH_{3}OH}}-\rm{PA_{\rm{out}}}|$   & 9    & 0.24 & 0.76      & 0.60 \\
$|\rm{PA}_{\rm{out}}-\langle\Phi_{\rm{B}}\rangle|$ & 9    & 0.39 & 1.23      & 0.10 \\
\\
\hline
\hline
\end{tabular}
\end{center}
\label{KS}
\end{table}\\
\indent On the contrary, if we consider the angle between $\rm{PA}_{\rm{out}}$ and $\langle\Phi_{\rm{B}}\rangle$ we find that the outflows are
primarily oriented along the magnetic fields. The corresponding CDF is shown in Fig.~\ref{cdf2} where, as in Fig.~\ref{cdf} we treat
 the full distributions, 
the dashed line is the CDF of a random distribution of the angles. The K-S test in this case gives a probability of about 10~\%. \\
\indent Although the sample is still small, this statistical analysis suggests that the magnetic fields close to the central YSO ($<1000$~AU)
are likely correlated with the direction of the large-scale outflows ($\sim1$~pc) detected in massive star-forming regions. At the same time any 
biases introduced in the magnetic field angles due to for example compression of maser gas are ruled out. To improve the statistics it is 
important to enlarge the number of sources of the magnetic field total sample for which the orientation of the molecular outflow is well
 determined. To do this, we are carrying out a large campaign to observe outflow tracers towards several massive star-forming regions.
\section{Conclusions}
We observed 4 massive star-forming regions at 6.7-GHz in full polarization spectral mode with the EVN to detect the linearly and circularly 
polarized emission of \meth ~masers. 
We detected a total of 128 6.7-GHz \meth ~masers and linearly polarized emission towards 24 of them. The linear polarization fraction is between
0.8~\% and 11.3~\%. 
We were also able to measure Zeeman-splitting in IRAS06058+2138-NIRS\,1 ($\Delta V_{\rm{Z}}=3.8\pm0.6~$\ms) and S255-IR 
($\Delta V_{\rm{Z}}=3.2\pm0.7~$\ms).\\
\indent We considered all the massive star-forming regions observed in full polarization mode at 6.7-GHz, so far, anywhere on the sky. By 
comparing the projected angles between magnetic fields and outflows, we find evidence that the magnetic field around massive YSOs are  
primarily oriented along the large-scale molecular outflows. This is supported by a Kolmogorov-Smirnov test that shows a 
probability of 10~\% that our distribution of angles is drawn from a random distribution. In the MFTS, the \meth ~masers are not primarily 
distributed along the outflows.\\
\indent Moreover, we empirically found that the linear polarization fraction of unsaturated \meth ~masers is $P_{\rm{l}}<4.5~\%$.
\appendix
\section{Tables}
In Tables~\ref{I06_tab}--\ref{S231_tab} we list all the \meth ~maser features detected towards the four massive star-forming regions
observed with the EVN. The Tables are organized as follows. The name of the feature is reported in Col.~1 and the group to which they belong 
is in Col.~2. The positions, Cols.~3 and 4, are referred to the maser feature used for self-calibration, no absolute positions were measured.
The peak flux density (I), the LSR velocity ($V_{\rm{lsr}}$), and the FWHM ($\Delta v\rm{_{L}}$) of the total intensity spectra of the 
maser features are reported in Cols.~5, 6, and 7, respectively. I, $V_{\rm{lsr}}$, and $\Delta v\rm{_{L}}$ are obtained using a Gaussian fit.
The linear polarization fraction ($P_{\rm{l}}$) and the linear polarization angles ($\chi$) are instead reported in Cols.~8 and 9, 
respectively. The best-fitting results obtained by using a model based on the radiative transfer theory of methanol masers 
for $\Gamma+\Gamma_{\nu}=1~\rm{s^{-1}}$ (Vlemmings et al. \cite{vle10}, Surcis et al. \cite{sur11b}) are reported in Cols.~10 (the emerging 
brightness temperature) and 11 (the intrinsic thermal linewidth). The errors were determined by analyzing the full probability distribution 
function. The angle between the magnetic field and the maser propagation direction ($\theta$, Col.~14) is determined by using the observed 
$P_{\rm{l}}$ and the fitted emerging brightness temperature. Also for $\theta$ the errors were determined by analyzing the full 
probability distribution function. The circular polarization fraction ($P_{\rm{V}}$) and the Zeeman-splitting ($\Delta V_{\rm{Z}}$) are finally
 in Cols.~12 and 13, respectively. The Zeeman-splitting is determined by fitting the V stoke spectra by using the best-fitting results ($\Delta 
V_{\rm{i}}$ and $T_{\rm{b}}\Delta\Omega$).
\begin {table*}[t!]
\caption []{All 6.7-GHz methanol maser features detected in IRAS06058+2138-NIRS\,1.} 
\begin{center}
\scriptsize
\begin{tabular}{ l c c c c c c c c c c c c c}
\hline
\hline
\,\,\,\,\,(1)&(2) & (3)       & (4)          & (5)            & (6)            & (7)                 & (8)         & (9)      & (10)                & (11)                     & (12)         & (13)                & (14)        \\
Maser & Group     & RA        & Dec          &Peak flux       & $V_{\rm{lsr}}$ & $\Delta v\rm{_{L}}$ &$P_{\rm{l}}$ &  $\chi$  & $\Delta V_{\rm{i}}$ & $T_{\rm{b}}\Delta\Omega$ & $P_{\rm{V}}$ & $\Delta V_{\rm{Z}}$ & $\theta$\\
      &       & offset        & offset       & Density(I)     &                &                     &             &	      &                     &                          &              &                     &           \\ 
      &       & (mas)         & (mas)        & (Jy/beam)      &  (km/s)        &      (km/s)         & (\%)        &   (\d)   & (km/s)              & (log K sr)               &  ($\%$)      & (m/s)               & (\d)       \\ 
\hline
IRAS06.01 &   B   & -101.002  & -242.539     & $0.11\pm0.01$  &     1.68       &  0.22               & $-$         & $-$      &  $-$                & $-$                      & $-$	      & $-$                 &$-$ \\ 
IRAS06.02 &   A   &  -56.857  &  -34.304     & $5.95\pm0.03$  &     11.63      &  0.32               & $-$         & $-$      &  $-$                & $-$                      & $-$	      & $-$          	    &$-$ \\ 
IRAS06.03 &   B   &  -53.453  & -252.033     & $0.18\pm0.01$  &     3.13       &  0.22               & $-$         & $-$      &  $-$                & $-$                      & $-$	      & $-$         	    &$-$ \\ 
IRAS06.04 &   A   &  -49.836  &  -29.320     & $0.11\pm0.01$  &     8.66       &  0.16               & $-$         & $-$      &  $-$                & $-$                      & $-$	      & $-$          	    &$-$ \\ 
IRAS06.05 &   A   &  -48.453  &  -15.698     & $1.09\pm0.03$  &     10.37      &  0.23               & $-$         & $-$      &  $-$                & $-$                      & $-$	      & $-$          	    &$-$ \\ 
IRAS06.06 &   A   &  -47.815  &   11.005     & $0.39\pm0.01$  &     8.71       &  0.28               & $-$         & $-$      &  $-$                & $-$                      & $-$	      & $-$          	    &$-$ \\ 
IRAS06.07 &   A   &  -46.485  &  -17.204     & $5.15\pm0.14$  &     11.03      &  0.15               & $3.2\pm0.1$ & $23\pm12$&  $-$                & $-$                      & $-$	      & $-$          	    & $-$ \\ 
IRAS06.08 &   A   &  -35.582  &  -31.330     & $3.43\pm0.01$  &     11.87      &  0.87               & $-$         & $-$      &  $-$                & $-$                      & $-$	      & $-$          	    &$-$ \\ 
IRAS06.09\tablefootmark{a}& A& -34.359 &-11.194 & $8.75\pm0.08$&     11.16     &  0.18               & $6.8\pm0.3$ & $76\pm6$ &  $0.6^{+0.1}_{-0.1}$& $9.8^{+0.1}_{-0.1}$      & $-$	      & $-$          	    &$90^{+13}_{-13}$ \\ 
IRAS06.10 &   A   &  -32.284  &  -26.491     & $0.41\pm0.02$  &     11.82      &  0.33               & $-$         & $-$      &  $-$                & $-$                      & $-$	      & $-$          	    &$-$ \\ 
IRAS06.11 &   A   &  -30.104  &  -11.690     & $22.21\pm0.36$ &     10.77      &  0.52               & $1.3\pm0.5$ & $69\pm10$&  $-$                & $-$                      & $-$          & $-$          	    &$-$ \\ 
IRAS06.12 &   A   &  -23.190  &  -22.720     & $2.32\pm0.03$  &     11.60      &  0.28               & $-$         & $-$      &  $-$                & $-$                      & $-$	      & $-$          	    &$-$ \\ 
IRAS06.13 &   A   &  -20.956  &    6.157     & $0.16\pm0.01$  &     8.40       &  0.15               & $-$         & $-$      &  $-$                & $-$                      & $-$	      & $-$          	    &$-$ \\ 
IRAS06.14 &   B   &  -20.424  & -240.074     & $0.09\pm0.002$ &     3.13       &  0.19               & $-$         & $-$      &  $-$                & $-$                      & $-$	      & $-$          	    &$-$ \\ 
IRAS06.15 &   A   &  -19.839  &   -0.356     & $2.15\pm0.06$  &     11.21      &  0.85               & $-$         & $-$      &  $-$                & $-$                      & $-$	      & $-$          	    & $-$ \\ 
IRAS06.16 &   A   &  -18.881  &   -6.853     & $0.64\pm0.02$  &     11.69      &  0.29               & $-$         & $-$      &  $-$                & $-$                      & $-$	      & $-$          	    &$-$ \\ 
IRAS06.17 &   A   &  -18.669  &   41.107     & $0.72\pm0.01$  &     8.53       &  0.19               & $-$         & $-$      &  $-$                & $-$                      & $-$	      & $-$          	    &$-$ \\ 
IRAS06.18 &   A   &  -14.733  &   22.949     & $0.14\pm0.01$  &     8.79       &  0.20               & $-$         & $-$      &  $-$                & $-$                      & $-$	      & $-$          	    &$-$ \\ 
IRAS06.19 &   A   &  -12.659  &   39.961     & $5.88\pm0.13$  &     11.12      &  0.28               & $-$         & $-$      &  $-$                & $-$                      & $-$	      & $-$          	    &$-$ \\ 
IRAS06.20 &   A   &  -11.223  &   80.557     & $0.25\pm0.01$  &     7.87       &  0.25               & $-$         & $-$      &  $-$                & $-$                      & $-$	      & $-$          	    &$-$ \\
IRAS06.21 &   A   &   -6.489  &   -1.888     & $2.51\pm0.03$  &     11.30      &  0.28               & $-$         & $-$      &  $-$                & $-$                      & $-$	      & $-$          	    &$-$ \\
IRAS06.22\tablefootmark{a}& A & 0 &    0     & $93.29\pm0.16$ &     10.86      &  0.28               & $6.0\pm1.3$ & $76\pm5$ &  $1.2^{+0.3}_{-0.5}$& $9.6^{+0.6}_{-0.4}$      & $0.3$        & $3.8\pm0.6$         &$73^{+17}_{-7} $ \\ 
IRAS06.23 &   A   &    1.117  &   43.013     & $10.27\pm0.29$ &     10.68      &  0.18               & $-$         & $-$      &  $-$                & $-$                      & $-$	      & $-$          	    &$-$ \\
IRAS06.24 &   A   &    1.649  &    4.763     & $20.64\pm0.23$ &     10.42      &  0.36               & $4.9\pm0.3$ & $49\pm2$ &  $-$                & $-$                      & $-$	      & $-$          	    &$-$ \\ 
IRAS06.25 &   A   &    2.819  &    2.865     & $1.17\pm0.03$  &     11.30      &  0.27               & $-$         & $-$      &  $-$                & $-$                      & $-$	      & $-$          	    &$-$ \\
IRAS06.26 &   A   &    6.542  &  -12.850     & $10.27\pm0.30$ &     10.68      &  0.13               & $9.2\pm4.2$ & $-84\pm4$&  $-$                & $-$                      & $-$	      & $-$          	    &$-$ \\ 
IRAS06.27 &   A   &   11.169  &  -40.705     & $19.49\pm0.23$ &     10.72      &  0.24               & $2.2\pm0.8$ & $61\pm6$ &  $-$                & $-$                      & $-$	      & $-$          	    & $-$ \\ 
IRAS06.28 &   A   &   15.371  &   11.868     & $5.41\pm0.17$  &     10.99      &  0.27               & $-$         & $-$      &  $-$                & $-$                      & $-$	      & $-$          	    &$-$ \\
IRAS06.29 &   A   &   15.743  &   58.292     & $0.20\pm0.01$  &     8.44       &  0.21               & $-$         & $-$      &  $-$                & $-$                      & $-$	      & $-$          	    &$-$ \\
IRAS06.30\tablefootmark{a}& A & 15.797 & 0.828 & $50.24\pm0.23$ &   10.59      &  0.21               & $2.3\pm0.2$ & $67\pm4$ &  $0.8^{+0.1}_{-0.3}$& $9.9^{+0.4}_{-0.1} $     & $-$	      & $-$          	    &$64^{+13}_{-38} $ \\ 
IRAS06.31 &   A   &   26.274  &   43.127     & $10.68\pm0.21$ &     10.33      &  0.15               & $-$         & $-$      &  $-$                & $-$                      & $-$	      & $-$          	    & $-$ \\
IRAS06.32 &   A   &   32.391  &    3.937     & $1.79\pm0.01$  &     9.98       &  0.15               & $-$         & $-$      &  $-$                & $-$                      & $-$	      & $-$          	    &$-$ \\
IRAS06.33 &   A   &   34.093  &   15.846     & $3.37\pm0.12$  &     10.29      &  0.13               & $-$         & $-$      &  $-$                & $-$                      & $-$	      & $-$          	    &$-$ \\
IRAS06.34 &   A   &   43.560  &    3.336     & $11.18\pm0.05$ &     9.76       &  0.22               & $2.4\pm0.1$ & $19\pm8$ &  $1.1^{+0.2}_{-0.3}$& $9.1^{+0.3}_{-0.1}$      & $-$	      & $-$          	    &$84^{+6}_{-41} $ \\ 
IRAS06.35 &   A   &   45.103  &  -24.460     & $7.13\pm0.22$  &     10.55      &  0.16               & $3.3\pm0.8$ & $69\pm10$&  $0.7^{+0.1}_{-0.2}$& $9.3^{+0.3}_{-0.7}$      & $-$          & $-$          	    &$84^{+6}_{-16} $ \\ 
IRAS06.36 &   A   &   47.496  &   -6.975     & $0.19\pm0.01$  &     9.41       &  0.35               & $-$         & $-$      &  $-$                & $-$                      & $-$	      & $-$          	    &$-$ \\
IRAS06.37 &   A   &   49.996  &   12.009     & $13.53\pm0.33$ &     10.55      &  0.15               & $1.9\pm0.4$ & $59\pm10$&  $0.8^{+0.2}_{-0.1}$& $8.9^{+0.3}_{-0.2}$      & $-$          & $-$          	    &$76^{+12}_{-37} $ \\ 
IRAS06.38 &   A   &   69.037  &   11.696     & $1.65\pm0.04$  &     9.67       &  0.22               & $-$         & $-$      &  $-$                & $-$                      & $-$	      & $-$          	    & $-$ \\
IRAS06.39 &   A   &   76.908  &   15.019     & $4.05\pm0.01$  &     9.76       &  0.22               & $-$         & $-$      &  $-$                & $-$                      & $-$	      & $-$          	    & $-$ \\
\hline
\end{tabular}
\end{center}
\tablefoot{
\tablefoottext{a}{Because of the degree of the saturation of these \meth ~masers $T_{\rm{b}}\Delta\Omega$ is underestimated, $\Delta V_{\rm{i}}$ 
and $\theta$ are overestimated.}
}
\label{I06_tab}
\end{table*}
\begin {table*}[t!]
\caption []{All 6.7-GHz methanol maser features detected in IRAS22272+6358A.} 
\begin{center}
\scriptsize
\begin{tabular}{ l c c c c c c c c c c c c c}
\hline
\hline
\,\,\,\,\,(1)&(2) & (3)      & (4)     & (5)             & (6)            & (7)                 & (8)         & (9)       & (10)                 & (11)                     & (12)         & (13)                & (14)       \\
Maser & Group     & RA       & Dec     &Peak flux        & $V_{\rm{lsr}}$ & $\Delta v\rm{_{L}}$ &$P_{\rm{l}}$ &  $\chi$   & $\Delta V_{\rm{i}}$  & $T_{\rm{b}}\Delta\Omega$ & $P_{\rm{V}}$ & $\Delta V_{\rm{Z}}$ & $\theta$\\
      &           & offset   & offset  & Density(I)      &                &                     &             &	          &                      &                          &              &                     &           \\ 
      &           & (mas)    & (mas)   & (Jy/beam)       &  (km/s)        &      (km/s)         & (\%)        &   (\d)    & (km/s)               & (log K sr)               &  ($\%$)      & (m/s)               & (\d)       \\ 
\hline
IRAS22.01 &       & -173.256 & 265.579 & $0.062\pm0.004$ & -10.51         &   0.23              & $-$         & $-$       &  $-$                 &  $-$                     & $-$          & $-$                 &  $-$       \\
IRAS22.02 &       & -138.277 & 144.333 & $0.171\pm0.005$ & -10.20         &   0.16              & $-$         & $-$       &  $-$                 &  $-$                     & $-$          & $-$                 &  $-$       \\
IRAS22.03 &       & -123.872 & 140.640 & $0.584\pm0.005$ & -10.15         &   0.23              & $-$         & $-$       &  $-$                 &  $-$                     & $-$          & $-$                 &  $-$       \\ 
IRAS22.04 &       & -115.587 & 130.711 & $1.378\pm0.005$ & -10.20         &   0.22              & $-$         & $-$       &  $-$                 &  $-$                     & $-$          & $-$                 &  $-$       \\  
IRAS22.05 &       & -101.605 & 128.204 & $0.094\pm0.006$ & -10.15         &   0.26              & $-$         & $-$       &  $-$                 &  $-$                     & $-$          & $-$                 &  $-$       \\  
IRAS22.06 &       & -97.550  & 130.402 & $0.098\pm0.008$ & -10.90         &   0.17              & $-$         & $-$       &  $-$                 &  $-$                     & $-$          & $-$                 &  $-$       \\
IRAS22.07 &       & -69.710  & 157.776 & $0.354\pm0.007$ & -10.86         &   0.24              & $-$         & $-$       &  $-$                 &  $-$                     & $-$          & $-$                 &  $-$       \\  
IRAS22.08 &       & -69.213  & 133.293 & $0.102\pm0.008$ & -10.90         &   0.19              & $-$         & $-$       &  $-$                 &  $-$                     & $-$          & $-$                 &  $-$       \\   
IRAS22.09 &       & -67.695  &  62.641 & $0.118\pm0.005$ & -9.93          &   0.14              & $-$         & $-$       &  $-$                 &  $-$                     & $-$          & $-$                 &  $-$       \\
IRAS22.10 &       & -65.755  & 179.375 & $0.081\pm0.005$ & -11.16         &   0.17              & $-$         & $-$       &  $-$                 &  $-$                     & $-$          & $-$                 &  $-$       \\
IRAS22.11 &       & -65.008  & 169.689 & $0.557\pm0.008$ & -10.94         &   0.20              & $-$         & $-$       &  $-$                 &  $-$                     & $-$          & $-$                 &  $-$       \\
IRAS22.12 &       & -64.784  & 168.640 & $0.467\pm0.008$ & -10.86         &   0.21              & $-$         & $-$       &  $-$                 &  $-$                     & $-$          & $-$                 &  $-$       \\
IRAS22.13 &       & -56.848  & 155.579 & $2.720\pm0.008$ & -10.90         &   0.21              & $0.8\pm0.2$ & $-75\pm5$ &  $1.1^{+0.1}_{-0.2}$ &  $8.5^{+0.4}_{-0.2}$     & $-$          & $-$                 &  $77^{+13}_{-38}$ \\  
IRAS22.14 &       & -47.295  & 140.812 & $1.989\pm0.008$ & -10.90         &   0.23              & $1.0\pm0.1$ & $68\pm17$ &  $1.2^{+0.1}_{-0.2}$ &  $8.6^{+0.3}_{-0.1}$     & $-$          & $-$                 &  $80^{+10}_{-38}$\\
IRAS22.15 &       & -44.210  & 129.570 & $0.310\pm0.008$ & -10.72         &   0.20              & $-$         & $-$       &  $-$                 &  $-$                     & $-$          & $-$                 &  $-$       \\ 
IRAS22.16 &       & -36.796  & 121.681 & $0.629\pm0.005$ & -10.72         &   0.17              & $-$         & $-$       &  $-$                 &  $-$                     & $-$          & $-$                 &  $-$       \\
IRAS22.17 &       & -35.900  & 130.512 & $1.245\pm0.007$ & -11.03         &   0.14              & $-$         & $-$       &  $-$                 &  $-$                     & $-$          & $-$                 &  $-$       \\
IRAS22.18 &       & -34.830  & 143.387 & $0.221\pm0.008$ & -10.94         &   0.14              & $-$         & $-$       &  $-$                 &  $-$                     & $-$          & $-$                 &  $-$       \\
IRAS22.19 &       & -29.780  & 96.726  & $2.658\pm0.007$ & -10.77         &   0.31              & $1.7\pm0.7$ & $-78\pm4$ &  $1.5^{+0.2}_{-0.2}$ &  $8.9^{+0.1}_{-1.0}$     & $-$          & $-$                 &  $90^{+54}_{-54}$\\
IRAS22.20 &       & -22.491  & 87.498  & $0.321\pm0.004$ & -11.30         &   0.22              & $-$         & $-$       &  $-$                 &  $-$                     & $-$          & $-$                 &  $-$       \\
IRAS22.21 &       & -13.783  & 47.733  & $0.220\pm0.006$ & -10.15         &   0.25              & $-$         & $-$       &  $-$                 &  $-$                     & $-$          & $-$                 &  $-$       \\
IRAS22.22 &       & 0        & 0       & $4.838\pm0.005$ & -12.83         &   0.30              & $-$         & $-$       &  $-$                 &  $-$                     & $-$          & $-$                 &  $-$       \\
IRAS22.23 &       & 0.373    & -10.288 & $0.750\pm0.004$ & -12.57         &   0.26              & $-$         & $-$       &  $-$                 &  $-$                     & $-$          & $-$                 &  $-$       \\
IRAS22.24 &       & 1.319    & 29.022  & $0.217\pm0.005$ & -12.31         &   0.17              & $-$         & $-$       &  $-$                 &  $-$                     & $-$          & $-$                 &  $-$       \\
IRAS22.25 &       & 47.668   & 216.942 & $0.117\pm0.008$ & -10.99         &   1.08              & $-$         & $-$       &  $-$                 &  $-$                     & $-$          & $-$                 &  $-$       \\
IRAS22.26 &       & 215.898  & 319.664 & $0.109\pm0.008$ & -10.99         &   0.15              & $-$         & $-$       &  $-$                 &  $-$                     & $-$          & $-$                 &  $-$       \\
\hline
\end{tabular}
\end{center}
\label{I22_tab}
\end{table*}
\begin {table*}[t!]
\caption []{All 6.7-GHz methanol maser features detected in S255-IR.} 
\begin{center}
\scriptsize
\begin{tabular}{ l c c c c c c c c c c c c c}
\hline
\hline
\,\,\,\,\,(1)&(2) & (3)  & (4)      & (5)             & (6)            & (7)                 & (8)         & (9)       & (10)                & (11)                     & (12)         & (13)                & (14)        \\
Maser & Group & RA       & Dec      & Peak flux       & $V_{\rm{lsr}}$ & $\Delta v\rm{_{L}}$ &$P_{\rm{l}}$ &  $\chi$   & $\Delta V_{\rm{i}}$ & $T_{\rm{b}}\Delta\Omega$ & $P_{\rm{V}}$ & $\Delta V_{\rm{Z}}$ & $\theta$\\
      &       & offset   & offset   & Density(I)      &                &                     &             &	       &                     &                          &              &                     &           \\ 
      &       & (mas)    & (mas)    & (Jy/beam)       &  (km/s)        &      (km/s)         & (\%)        &   (\d)    & (km/s)              & (log K sr)               &  ($\%$)      & (m/s)               & (\d)       \\ 
\hline 
S255.01 & A   & -217.801 & 97.237   & $0.209\pm0.009$ &    4.96        &      $0.13$         & $-$         & $-$       &  $-$                & $-$                      & $-$          & $-$                 &$-$ \\ 
S255.02 & B   & -217.039 & -41.428  & $0.156\pm0.003$ &    5.70        &      $0.13$         & $-$         & $-$       &  $-$                & $-$                      & $-$          & $-$          	     &$-$ \\ 
S255.03 & B   & -212.685 & -62.891  & $1.212\pm0.005$ &    5.48        &      $0.13$         & $-$         & $-$       &  $-$                & $-$                      & $-$          & $-$          	     &$-$ \\ 
S255.04 & B   & -212.522 & -124.538 & $3.113\pm0.015$ &    4.87        &      $0.17$         & $2.2\pm0.5$ & $39\pm2$  &  $0.8^{+0.2}_{-0.1}$& $9.0^{+0.3}_{-0.4}$      & $-$          & $-$          	     &$80^{+10}_{-21}$ \\ 
S255.05 & B   & -212.413 & -109.398 & $0.501\pm0.007$ &    5.00        &      $0.21$         & $-$         & $-$       &  $-$                & $-$                      & $-$          & $-$          	     &$-$ \\ 
S255.06 & B   & -206.100 & -53.278  & $1.707\pm0.005$ &    5.57        &      $0.17$         & $3.8\pm0.2$ & $39\pm4$  &  $0.7^{+0.1}_{-0.1}$& $9.3^{+0.1}_{-0.1}$      & $-$	       & $-$          	     &$85^{+4}_{-10}$ \\ 
S255.07 & B   & -205.610 & -86.506  & $0.343\pm0.004$ &    5.22        &      $0.16$         & $-$         & $-$       &  $-$                & $-$                      & $-$	       & $-$          	     &$-$ \\ 
S255.08 & B   & -204.522 & -132.717 & $2.343\pm0.018$ &    4.74        &      $0.16$         & $1.0\pm0.3$ & $60\pm11$ &  $0.8^{+0.1}_{-0.1}$& $8.6^{+0.4}_{-0.4}$      & $-$	       & $-$          	     &$74^{+14}_{-38}$ \\ 
S255.09 & B   & -203.923 & -19.556  & $0.399\pm0.006$ &    5.44        &      $0.14$         & $-$         & $-$       &  $-$                & $-$                      & $-$	       & $-$          	     &$-$ \\ 
S255.10 & B   & -201.583 & -70.501  & $0.452\pm0.005$ &    5.35        &      $0.14$         & $-$         & $-$       &  $-$                & $-$                      & $-$	       & $-$          	     &$-$ \\ 
S255.11 & B   & -199.624 & -87.763  & $0.686\pm0.014$ &    4.78        &      $0.14$         & $-$         & $-$       &  $-$                & $-$                      & $-$	       & $-$          	     &$-$ \\ 
S255.12 & B   & -195.977 & -140.797 & $0.297\pm0.011$ &    4.47        &      $0.15$         & $-$         & $-$       &  $-$                & $-$                      & $-$	       & $-$          	     &$-$ \\ 
S255.13 & B   & -195.488 & -104.160 & $0.270\pm0.004$ &    5.40        &      $0.13$         & $-$         & $-$       &  $-$                & $-$                      & $-$	       & $-$          	     &$-$ \\ 
S255.14 & B   & -193.800 & -58.836  & $0.256\pm0.005$ &    5.40        &      $0.63$         & $-$         & $-$       &  $-$                & $-$                      & $-$	       & $-$          	     &$-$ \\ 
S255.15 & B   & -187.760 & -151.674 & $0.198\pm0.004$ &    4.25        &      $0.84$         & $-$         & $-$       &  $-$                & $-$                      & $-$	       & $-$          	     &$-$ \\ 
S255.16 & B   & -172.086 & -97.004  & $0.176\pm0.006$ &    4.39        &      $1.29$         & $-$         & $-$       &  $-$                & $-$                      & $-$	       & $-$          	     &$-$ \\ 
S255.17 & B   & -164.521 & -108.042 & $0.250\pm0.004$ &    4.08        &      $0.20$         & $-$         & $-$       &  $-$                & $-$                      & $-$	       & $-$          	     &$-$ \\ 
S255.18 & B   & -162.126 & -156.319 & $0.394\pm0.003$ &    4.12        &      $0.22$         & $-$         & $-$       &  $-$                & $-$                      & $-$	       & $-$          	     &$-$ \\ 
S255.19 & B   & -160.276 & -155.373 & $0.334\pm0.003$ &    4.08        &      $0.24$         & $-$         & $-$       &  $-$                & $-$                      & $-$	       & $-$          	     &$-$ \\ 
S255.20 & B   & -158.534 & -188.637 & $0.116\pm0.004$ &    4.17        &      $0.20$         & $-$         & $-$       &  $-$                & $-$                      & $-$	       & $-$          	     &$-$ \\ 
S255.21 & B   & -148.194 & -146.700 & $0.668\pm0.003$ &    4.03        &      $0.22$         & $-$         & $-$       &  $-$                & $-$                      & $-$	       & $-$          	     &$-$ \\ 
S255.22 & B   & -144.548 & -138.336 & $0.259\pm0.004$ &    4.17        &      $0.17$         & $-$         & $-$       &  $-$                & $-$                      & $-$	       & $-$          	     &$-$ \\ 
S255.23 & A   & -142.207 & 231.489  & $0.268\pm0.016$ &    4.82        &      $0.15$         & $-$         & $-$       &  $-$                & $-$                      & $-$	       & $-$          	     &$-$ \\ 
S255.24 & C   & -135.513 & -399.889 & $0.130\pm0.003$ &    3.38        &      $0.18$         & $-$         & $-$       &  $-$                & $-$                      & $-$	       & $-$          	     &$-$ \\ 
S255.25 & C   & -128.166 & -393.681 & $0.123\pm0.004$ &    2.23        &      $0.27$         & $-$         & $-$       &  $-$                & $-$                      & $-$	       & $-$          	     &$-$ \\ 
S255.26 & C   & -125.989 & -373.406 & $0.361\pm0.003$ &    3.03        &      $0.23$         & $-$         & $-$       &  $-$                & $-$                      & $-$	       & $-$          	     &$-$ \\ 
S255.27 & C   & -125.935 & -361.589 & $0.180\pm0.003$ &    2.28        &      $0.34$         & $-$         & $-$       &  $-$                & $-$                      & $-$	       & $-$          	     &$-$ \\ 
S255.28 & C   & -120.166 & -351.229 & $0.104\pm0.004$ &    1.31        &      $0.23$         & $-$         & $-$       &  $-$                & $-$                      & $-$	       & $-$          	     &$-$ \\ 
S255.29 & A   & -69.172  & 409.050  & $0.280\pm0.016$ &    4.82        &      $0.68$         & $-$         & $-$       &  $-$                & $-$                      & $-$	       & $-$          	     &$-$ \\ 
S255.30\tablefootmark{a}& B & 0 & 0 & $10.636\pm0.015$&    4.61        &      $0.24$         & $4.5\pm0.3$ & $23\pm5$  &  $1.1^{+0.1}_{-0.4}$& $9.5^{+0.3}_{-0.1}$      & $0.3$	       & $3.2\pm0.7$         &$82^{+8}_{-18}$ \\ 
S255.31 & B   & 6.694    & 5.865    & $3.689\pm0.014$ &    4.78        &      $0.20$         & $1.8\pm0.4$ & $21\pm7$  &  $0.9^{+0.1}_{-0.2}$& $9.0^{+0.4}_{-0.3}$      & $-$	       & $-$          	     &$79^{+11}_{-36}$ \\ 
\hline
\end{tabular}
\end{center}
\tablefoot{
\tablefoottext{a}{Because of the degree of the saturation of these \meth ~masers $T_{\rm{b}}\Delta\Omega$ is underestimated, $\Delta V_{\rm{i}}$ 
and $\theta$ are overestimated.}
}
\label{S255_tab}
\end{table*}
\begin {table*}[t!]
\caption []{All 6.7-GHz methanol maser features detected in S231.} 
\begin{center}
\scriptsize
\begin{tabular}{ l c c c c c c c c c c c c c}
\hline
\hline
\,\,\,\,\,(1)&(2)& (3)     & (4)      & (5)             & (6)            & (7)                 & (8)         & (9)       & (10)                & (11)                     & (12)         & (13)                & (14)        \\
Maser & Group   & RA       & Dec      &Peak flux        & $V_{\rm{lsr}}$ & $\Delta v\rm{_{L}}$ &$P_{\rm{l}}$ &  $\chi$   & $\Delta V_{\rm{i}}$ & $T_{\rm{b}}\Delta\Omega$ & $P_{\rm{V}}$ & $\Delta V_{\rm{Z}}$ & $\theta$\\
      &         & offset   & offset   & Density(I)      &                &                     &             &	         &                     &                          &              &                     &           \\ 
      &         & (mas)    & (mas)    & (Jy/beam)       &  (km/s)        &      (km/s)         & (\%)        &   (\d)    & (km/s)              & (log K sr)               &  ($\%$)      & (m/s)               & (\d)       \\ 
\hline 
S231.01 &   C   & -108.670 & 51.949   & $2.287\pm0.003$ &   -11.82       &  0.42               & $0.8\pm0.4$ & $-7\pm5$  &  $1.9^{+0.1}_{-0.2}$& $8.5^{+0.9}_{-0.1}$      & $-$          & $-$                 &$90^{+60}_{-60}$ \\ 
S231.02 &   B   & -68.055  & 47.474   & $0.177\pm0.003$ &   -11.51       &  0.22               &    $-$      & $-$       &  $-$                & $-$       		  & $-$          & $-$                 &$-$ \\ 
S231.03 &   B   & -54.975  & 34.267   & $0.224\pm0.003$ &   -11.20       &  0.22               &    $-$      & $-$       &  $-$                & $-$                      & $-$          & $-$                 &$-$ \\ 
S231.04 &   B   & -53.140  & 70.908   & $0.053\pm0.003$ &   -11.29       &  0.14               &    $-$      & $-$       &  $-$                & $-$                      & $-$          & $-$                 &$-$ \\ 
S231.05 &   B   & -50.597  & -3.765   & $0.053\pm0.003$ &   -11.29       &  0.18               &    $-$      & $-$       &  $-$                & $-$                      & $-$          & $-$                 &$-$ \\ 
S231.06 &A$_{2}$& -48.914  & -104.523 & $0.062\pm0.003$ &   -14.27       &  0.19               &    $-$      & $-$       &  $-$                & $-$                      & $-$	         & $-$                 &$-$ \\ 
S231.07 &   B   & -47.498  & 32.341   & $0.096\pm0.003$ &   -11.20       &  0.15               &    $-$      & $-$       &  $-$                & $-$                      & $-$	         & $-$                 &$-$ \\ 
S231.08 &   B   & -46.860  & 45.284   & $0.107\pm0.003$ &   -11.86       &  0.23               &    $-$      & $-$       &  $-$                & $-$                      & $-$	         & $-$                 &$-$ \\ 
S231.09 &   B   & -46.790  & 33.001   & $0.073\pm0.003$ &   -11.33       &  0.22               &    $-$      & $-$       &  $-$                & $-$                      & $-$	         & $-$                 &$-$ \\ 
S231.10 &A$_{2}$& -45.954  & -87.692  & $0.310\pm0.004$ &   -14.01       &  0.29               &    $-$      & $-$       &  $-$                & $-$                      & $-$	         & $-$                 &$-$ \\ 
S231.11\tablefootmark{a} &A$_{2}$& -42.031  & -74.463& $2.472\pm0.004$ &   -13.79 &  0.32      & $11.3\pm0.6$& $-74\pm2$ &  $<0.5$             & $11.3^{+0.3}_{-0.1}$     & $-$	         & $-$                 &$82^{+8}_{-18}$ \\ 
S231.12 &A$_{2}$& -40.418  & -90.500  & $1.895\pm0.003$ &   -6.37        &  0.32               &    $-$      & $-$       &  $-$                & $-$                      & $-$	         & $-$                 &$-$ \\ 
S231.13 &  B    & -38.131  & 21.629   & $0.065\pm0.003$ &   -11.20       &  0.19               &    $-$      & $-$       &  $-$                & $-$                      & $-$	         & $-$                 &$-$ \\ 
S231.14 &A$_{2}$& -32.884  & -82.264  & $0.156\pm0.003$ &   -7.69        &  0.36               &    $-$      & $-$       &  $-$                & $-$                      & $-$	         & $-$                 &$-$ \\ 
S231.15 &  B    & -27.870  & 30.174   & $0.799\pm0.003$ &   -11.46       &  0.21               &    $-$      & $-$       &  $-$                & $-$                      & $-$	         & $-$                 &$-$ \\ 
S231.16 &A$_{1}$& -21.973  & -36.274  & $0.257\pm0.003$ &   -15.02       &  0.39               &    $-$      & $-$       &  $-$                & $-$                      & $-$	         & $-$                 &$-$ \\ 
S231.17 &A$_{1}$&  -19.071 & -30.800  & $0.620\pm0.003$ &   -14.23       &  0.33               &    $-$      & $-$       &  $-$                & $-$                      & $-$	         & $-$                 &$-$ \\ 
S231.18 &A$_{1}$& -16.947  & -23.762  & $0.143\pm0.003$ &   -14.54       &  0.27               &    $-$      & $-$       &  $-$                & $-$                      & $-$	         & $-$                 &$-$ \\ 
S231.19 &A$_{1}$& -9.147   & 11.021   & $0.174\pm0.006$ &   -13.18       &  0.42               &    $-$      & $-$       &  $-$                & $-$                      & $-$	         & $-$                 &$-$ \\ 
S231.20\tablefootmark{a} &A$_{1}$& -7.882 & -11.410  & $3.900\pm0.004$ &   -14.14  &  0.34     & $5.8\pm1.1$ & $47\pm1$  &  $1.2^{+0.3}_{-0.5}$& $9.6^{+0.5}_{-0.7}$      & $-$	         & $-$                 &$83^{+7}_{-16}$ \\ 
S231.21 &A$_{1}$& -3.471   & -13.863  & $0.261\pm0.008$ &   -13.13       &  0.33               &    $-$      & $-$       &  $-$                & $-$                      & $-$	         & $-$                 &$-$ \\ 
S231.22 &A$_{1}$&  -2.403  & -4.982   & $0.539\pm0.004$ &   -13.66       &  0.39               &    $-$      & $-$       &  $-$                & $-$                      & $-$	         & $-$                 &$-$ \\ 
S231.23 &A$_{1}$& 0        & 0        & $23.419\pm0.010$&   -12.96       &  0.34               & $4.0\pm0.4$ & $48\pm1$  &  $1.7^{+0.4}_{-0.5}$& $9.3^{+0.4}_{-0.2}$      & $-$	         & $-$                 &$80^{+8}_{-41}$ \\ 
S231.24 &A$_{1}$& 5.467    & 9.663    & $0.612\pm0.005$ &   -13.31       &  0.38               &    $-$      & $-$       &  $-$                & $-$                      & $-$	         & $-$                 &$-$ \\ 
S231.25\tablefootmark{a} &A$_{1}$& 5.815 & 4.520     & $6.634\pm0.011$ &   -12.87  &  0.30     & $5.3\pm0.6$ & $48\pm1$  &  $1.3^{+0.2}_{-0.5}$& $9.6^{+0.4}_{-0.1}$      & $-$	         & $-$                 &$82^{+8}_{-18}$ \\ 
S231.26 &   D   & 14.231   & -108.620 & $0.105\pm0.003$ &   -12.17       &  0.22               &    $-$      & $-$       &  $-$                & $-$                      & $-$	         & $-$                 &$-$ \\ 
S231.27 &A$_{1}$& 15.206   & 18.288   & $0.389\pm0.005$ &   -13.22       &  0.34               &    $-$      & $-$       &  $-$                & $-$                      & $-$	         & $-$                 &$-$ \\ 
S231.28 &A$_{1}$& 17.330   & 21.011   & $0.153\pm0.004$ &   -13.35       &  0.40               &    $-$      & $-$       &  $-$                & $-$                      & $-$	         & $-$                 &$-$ \\ 
S231.29 &   D   & 22.588   & -116.097 & $0.294\pm0.003$ &   -12.08       &  0.25               &    $-$      & $-$       &  $-$                & $-$                      & $-$	         & $-$                 &$-$ \\ 
S231.30 &   D   & 25.038   & -109.062 & $0.302\pm0.010$ &   -12.96       &  0.15               &    $-$      & $-$       &  $-$                & $-$                      & $-$	         & $-$      	       &$-$ \\ 
S231.31 &   D   & 28.148   & -117.458 & $0.188\pm0.006$ &   -13.18       &  0.90               &    $-$      & $-$       &  $-$                & $-$                      & $-$	         & $-$                 &$-$ \\ 
S231.32 &   D   & 28.949   & -67.051  & $0.103\pm0.003$ &   -12.39       &  0.18               &    $-$      & $-$       &  $-$                & $-$                      & $-$	         & $-$                 &$-$ \\ 
\hline
\end{tabular}
\end{center}
\tablefoot{
\tablefoottext{a}{Because of the degree of the saturation of these \meth ~masers $T_{\rm{b}}\Delta\Omega$ is underestimated, $\Delta V_{\rm{i}}$ 
and $\theta$ are overestimated.}
}
\label{S231_tab}
\end{table*}\\
%

\noindent \small{\textit{Acknowledgements.} We wish to thank the referee Dr.~A. Sanna for making useful suggestions that have improved the paper.
This research was partly supported by the Deutsche Forschungsgemeinschaft (DFG) through the Emmy Noether 
Research grant VL 61/3--1. L.H.Q.-N. thanks the JIVE Summer Student Programme 2012. The EVN is a joint facility of European, Chinese, South 
African, and other radio astronomy institutes funded by their national research councils. }

\end{document}